\documentclass[%
  fontsize=10pt, 
  BCOR=0cm,
  DIV=9,
  abstract=true,
  twocolumn=false,
  titlepage=false
]{scrartcl}

\usepackage{ifluatex}
\ifluatex%
    \usepackage[default]{fontsetup}
\else
    \usepackage[utf8]{inputenc}
    \usepackage[T1]{fontenc}
    \usepackage{amssymb}
\fi

\usepackage{scrlayer-scrpage}
\usepackage{csquotes}


\usepackage{graphicx}
\usepackage{tikz}
\usetikzlibrary{arrows.meta,positioning, arrows.meta, decorations.pathmorphing}

\usepackage[style=ieee,backend=biber]{biblatex}
\usepackage[ngerman, english]{babel}
\usepackage{microtype}
\usepackage{hyperref}
\usepackage{orcidlink}  
\hypersetup{%
  colorlinks      = {false},
  pdfborder       = {0 0 0},
  pdftitle        = {Computability of the Optimizer for Rate Distortion Functions},
  pdfauthor       = {Jonathan E. W. Huffmann},
  pdfkeywords     = {Rate distortion theory, source coding, Turing computability, computability, 
  computation of optimal channels},
}
\usepackage[labelfont=bf]{caption}

\usepackage{amsthm,amsmath}
\newtheorem{theorem}{Theorem}
\newtheorem{definition}{Definition}

\newtheorem{lemma}{Lemma}

\newtheorem{corollary}{Corollary}

\newtheorem{remark}{Remark}

\usepackage{scrlayer-scrpage}
\setkomafont{title}{\bfseries\rmfamily\scshape}
\setkomafont{author}{\bfseries\small}
\setkomafont{date}{\small}
\setkomafont{sectioning}{\rmfamily\bfseries\scshape}
\setkomafont{section}{\centering\normalsize}
\setkomafont{subsection}{\centering\normalsize}
\setkomafont{subsubsection}{\centering\normalsize}

\setcapindent{0pt} 
\setkomafont{captionlabel}{\bfseries} 

\makeatletter
\def\@seccntformat#1{\@ifundefined{#1@cntformat}%
    {\csname the#1\endcsname\space}
    {\csname #1@cntformat\endcsname}}
\newcommand\section@cntformat{\thesection\@.\space} 
\makeatother
\renewcommand\thesection{\Roman{section}}

\usepackage{xpatch}
\makeatletter%
  \xpatchcmd{\maketitle}{\usekomafont{title}{\huge \@title\par}}%
            {\usekomafont{title}{\large \@title\par}}{}{}
  \xpatchcmd{\@maketitle}{\usekomafont{title}{\huge \@title\par}}%
            {\usekomafont{title}{\large \@title\par}}{}{}
\makeatother%

\title{Computability of the Optimizer for Rate Distortion Functions}
\author{Jonathan E. W. Huffmann\,\orcidlink{0000-0003-2465-7181}\thanks{Jonathan E. W. Huffmann is with the Chair of Theoretical Information Technology, Technical University of Munich, Germany~(email: \href{mailto:jonathan.huffmann@tum.de}{\texttt{jonathan.huffmann@tum.de}})}
\and%
  Holger Boche\,\orcidlink{0000-0002-8375-8946}\thanks{Holger Boche is with the Chair of Theoretical Information Technology, Technical University of Munich,%
  BMFTR Research Hub 6G-life, Germany,%
  Munich Center for Quantum Science and Technology (MCQST), Germany,%
  and also with the Munich Quantum Valley (MQV), Germany~(email: \href{mailto:boche@tum.de}{\texttt{boche@tum.de}})}
}
\date{\today}

\begin{document}
\maketitle
\begin{abstract}
  Rate distortion theory treats the problem of encoding a source with minimum
  codebook size while at the same time allowing for a certain amount of errors
  in the reconstruction measured by a fidelity criterion and distortion
  level. Similar to the channel coding problem the optimal rate of the codebook
  with respect to the blocklength is given by a convex optimization problem
  involving information theoretic quantities like mutual information. The value
  of the rate in dependence of the distortion level as well as the optimizer
  used in the codebook construction are of theoretical and practical importance
  in communication and information theory.

  In this paper the behavior of the rate distortion function regarding the
  computability of the optimizing test channel is investigated. We find that
  comparable with known results about the optimizer for other information
  theoretic problems a similar result is found to be true also regarding the
  computability of the optimizer for rate distortion functions.

  It turns out that while the rate distortion function is usually computable the
  optimizer for this problem is in general non-computable even for simple
  distortion measures.
\end{abstract}

\section{Introduction}
Rate distortion theory is a notion of source coding where instead of providing a
one-to-one mapping between source sequences and codebook with asymptotically
vanishing approximation error the source is encoded with a fidelity criterion
allowing for some controllable error to be made in the reconstruction. This
allows encoding of discrete as well as continuous sources which otherwise could
not be encoded with finite or otherwise limited coding rates as their discrete
entropy is for example not a finite quantity or too large and therefore the
standard procedures used in source and channel coding do not work as
required. rate distortion theory was first introduced by Shannon in his seminal
paper~\cite{Shannon1948}.

The rate distortion function describes the minimum exponential rate of a
codebook in terms of a variable reconstruction error which is defined by a
distortion measure. The rate distortion function is therefore a function of the
source probability a distortion measure and some variable distortion
level. 

Similar to the channel capacity the rate distortion function can be shown to be
given by a convex optimization problem of mutual information. In contrast to
channel coding the optimization is over the transition probability while the
source probability is assumed to be known. The transition probabilities in this
setting are also commonly referred to as test channels.\cite{Gallager1968}

The problem of finding the rate distortion function as well as the optimizer and
appropriate coding schemes are not that exhaustively treated in the
literature. Most textbooks on information theory treat the channel coding
problem in great detail while only dedicating a small portion with only some
repeating examples on rate distortion theory. Moreover even for these cases no
closed form solution exists for the optimal test channel probabilities.

Despite that source coding with a fidelity criterion is of great theoretical and
practical interest as continuous sources and lossy compression are ubiquitous
present in real world phenomenon. This includes applications like compressed
sensing, lossy compression, and joint sensing and communication as are also
commonly found in proposed 6G applications~\cite{Fettweis2021}.

A treatment of rate distortion theory for the finite blocklength regime is found
in~\cite{Kostina2012,Kostina2014}. Providing some more insights into the
non-asymptotic behaviour of the rate distortion function.

Nowadays digital computers are more and more used to simulate and approximate
the performance metrics of various communication systems and models.

Because the analytic computation of the rate distortion function and the
optimizing transition probability matrix is often challenging even for simple
examples like the Hamming-distortion measure, numerical calculations and
algorithms are typically employed to gain insight into these problems.

An algorithm for the computation of the channel capacity was first proposed by
Meister~\cite{Meister1967} for some special cases. Later
Blahut~\cite{Blahut1972} and Arimoto~\cite{Arimoto1972} proposed a general
algorithm for computing the channel capacity. Using parallels of the calculation
of channel capacity Blahut~\cite{Blahut1972} also extended his algorithm to rate
distortion problems. While his proof was later extended by~\cite{Csiszar1974}
who also pointed out that for the general problem no unique optimizing
distribution exists. This limits the existence of an universally stopping
criterion for the optimizer as will be shown in this paper. Despite that
Arimoto~\cite{Arimoto1972} gives an error estimate also for the capacity
achieving input distribution under certain conditions on the channel matrix. An
equivalent result for rate distortion functions is found
in~\cite{Boukris1973}. As it will turn out the conditions under which these
bounds hold cannot be checked by a Turing machine and are therefore not
algorithmically decidable in general. Later the Blahut-Arimoto type of algorithms
were even further generalized in~\cite{Csiszar1984} by Csizár and Tusnády to a
broader class of problems using methodologies of information geometry.

The high availability of digital computers have led to a high number of
practical examples and extensions of the Blahut-Arimoto type algorithm to
various channel coding and rate distortion problems
(see~\cite{Ugur2017,Dupuis2004,Vontobel2008, Naiss2013}).There exist also
extensions of the Blahut-Arimoto algorithms for classical quantum channels
(cf.~\cite{Li2019}).

While the Blahut-Arimoto algorithms~\cite{Blahut1972,Arimoto1972} give a way of
computing channel capacity and the rate distortion function algorithmically,
analytical solutions to the rate distortion function and its optimizer are in
general not known and often no closed form solutions exist
(see~\cite{Berger1971}). The Blahut-Arimoto type algorithms can in theory be
used to also compute the optimizer in some well behaved special cases.  As there
still exist no general algorithms to directly compute an optimizer effectively
in these cases whether there exist computable algorithms at all is of high
interest.

Despite the advance of modern digital computers more and more problems in
communication and information theory have been shown to be not generally or only
partly Turing
computable~\cite{Grigorescu2025,Grigorescu2024a,Boche2019,Boche2020a}. This is
especially true for the underlying algorithms and optimization problems used in
many communication problems like convex optimization~\cite{Grigorescu2024} and
spectral factorization~\cite{Boche2020}.
  
In~\cite{Boche2023} the authors showed that the problem of finding the
optimizing input distribution of the channel coding problem is in general not
Turing computable.  In this Paper we investigate the computability of the
optimizer for rate distortion problems. We show that a similar behaviour is
found to be true for the rate distortion function. Further results regarding
computability of optimizer in classical information theoretic problems as well
as in a more general setting are found in~\cite{Lee2024}. Nevertheless without
treating the rate distortion problem explicitly.

The proof ideas used in\cite{Boche2023} to show that the capacity achieving
input distribution is not generally Turing computable rely heavily on simple
properties of mutual information. For the rate distortion function the
dependence on mutual information is more complex and complicated by the
additional dependence on a specific distortion measure and distortion level.
  
The ideas used in~\cite{Boche2023,Lee2024} therefore do not work in this case. The
question of the computability of the optimizer for rate distortion functions
has since then been an interesting open problem~\cite{Stylianou2024}.

A Turing machine is a mathematical model for an idealized computation
machine. This model was initially used by Turing to proof that there exist
non-computable real numbers and to solve the famous decision problem in
logic.\cite{Turing1937,Goedel1930}
  
These are typically defined to have an input and output tape with separate
fields containing only blanks, zeros and ones. A finite number of registers and
logic are then used to manipulate the input and output tape. The amount of input
and output tape, registers and logic are not limited, thus Turing machines have
no limitation on complexity of the functions which can be calculated providing a
theoretical model to describe arbitrary complex algorithms and computing
machines. The Turing machine thus provides the blueprint for all digital
computers used today.

In a lot of optimization problems the extremum of a sufficiently well
behaved and understood function $f:\mathbb{R}^{n}\to\mathbb{R}$ is sought.
\begin{equation}
    f_{\min}=\min_{x\in\mathbb{R}}f(x)
\end{equation}
In such cases the minimum or maximum value of $f$ can often be shown to be
computable under mild conditions on $f$. 
Nevertheless an important question which is of utmost practical importance is
for the computability of the optimizer $x^{*}$.
\begin{equation}
  x^{*}=\mathrm{argmax}\left(f(x)\right)
\end{equation}
It was shown by Specker in~\cite{Specker1959} that the optimizer $x^{*}$ in
these cases need not be computable even if $f$ is.

In this paper we analyze the computability of the optimizing test channel
probability for rate distortion functions. For this we employ the notion of
recursive functions. Recursive functions were shown to be exactly the functions
which can be calculated on a Turing machine and thus on any digital computer. 

As it turns out the general problem of computing the optimizer is not a Turing
computable problem and therefore not feasible even on modern high speed and high
memory digital computers.

\section{Notation}
By $\mathbb{N}$, $\mathbb{Q}$, $\mathbb{R}$ we denote the non-negative integers,
the rational numbers and the real numbers respectively. We $\bar{\mathbb{R}}$
write for the extended real numbers and $\mathbb{R}_{\geq0}$ for the
non-negative real numbers. For $n$-dimensional vectors
\(x^{(n)}\in\mathbb{R}^{n}\) we write \(x^{(n)}=(x_{0},x_{1},\dots,x_{n-1})\) as
is often done in the information theoretic literature.  For matrices
\(A\in\mathbb{R}^{M\times N}\) we write \({(a_{i,j})}_{0\leq i\leq M-1,0\leq
  j\leq N-1}\). We further define p-norms for matrices by
\begin{equation}
  \|A\|_{p}={\left(\sum_{i=0}^{M-1}\sum_{j=0}^{N-1}|a_{i,j}|^{p}\right)}^{\frac{1}{p}}.
\end{equation}
This leads to the important case
\begin{equation}
  \|A\|_{2}=\sqrt{\sum_{i=0}^{M-1}\sum_{j=0}^{N-1}|a_{i,j}|^{2}}
\end{equation}
for $p=2$. Sets are denoted by $\mathcal{X}$ and $\mathcal{Y}$. By
$|\mathcal{X}|$ we denote the size of the set $\mathcal{X}$. Without loss of
generality and when the elements of a set are not important we typically denote
the elements of $\mathcal{X}$ by a subset of the natural numbers to simplify
notation. In the context of information theory and especially source coding
these sets are commonly referred to as alphabets.

Probability distributions are denoted by $P_{X}$, $P_{Y}$ as well as $P_{Y|X}$
for the conditional probability distribution. Probability distributions are
assumed to be defined by appropriate sigma algebras and product sigma algebras
on the given sets.

For two discrete conditional probability distributions $P_{Y|X}$ and $Q_{Y|X}$
we define the total variation distance between the distributions by
\begin{equation}
  \|P_{Y|X}-Q_{Y|X}\|_{TV}=\max_{x\in\mathcal{X}}
  \sum_{y\in\mathcal{Y}}|P_{Y|X}(y|x)-P_{Y|X}(y|x)|.
\end{equation}

The support of a discrete probability distribution is 
\begin{equation}
  \mathrm{supp}(P_{X}):=\left\{x\in\mathcal{X}|P_{X}(x)\neq0\right\}
\end{equation}
the subset, with nonzero probability, of the set it is defined on.

\section{Prerequisites from Information Theory}
In the following we shortly define and introduce the standard measures of
information theory.  

Mutual information between two random variables $X$ and $Y$ defined on the sets
$\mathcal{X}$ and $\mathcal{Y}$ respectively is given by
\begin{equation}
  I(X;Y)=\sum_{x\in\mathcal{X}}
  \sum_{y\in\mathcal{Y}}P_{X,Y}(x,y)\log\left(\frac{P_{X,Y}(x,y)}{P_{X}(x)P_{Y}(y)}\right).
\end{equation}

The entropy of a random variable $X$ on the alphabet $\mathcal{X}$ is defined by
\begin{equation}
  H(X) = -\sum_{x\in\mathcal{X}}P_{X}(x)\log\left(P_{X}(x)\right).
\end{equation}

The Conditional entropy of a random variable $X$ given a value of the random
variable $Y=y$ is given by
\begin{equation}
  H(X|Y=y)=-\sum_{x\in\mathcal{X}}P_{X|Y}(x|y)\log\left(P_{X|Y}(x|y)\right).
\end{equation}

This definition leads to the conditional entropy of $X$ given $Y$ by averaging
over all values of $y$.
\begin{equation}
  H(X|Y)=\sum_{y\in\mathcal{Y}}P_{Y}(y)H(X|Y=y)=
  -\sum_{x\in\mathcal{X}}\sum_{y\in\mathcal{Y}}P_{X|Y}(x|y)P_{Y}(y)\log\left(P_{X|Y}(x|y)\right)
\end{equation}

On the set of $n$-dimensional probability vectors $\mathcal{P}(\mathcal{X}^{n})$
over the alphabet $\mathcal{X}$ we define the following partial order
\begin{equation}
  P_{X}\prec
  P_{Y}\quad\iff\quad\sum_{k=0}^{l}P^{\downarrow}_{X}(k)\leq\sum_{k=0}^{l}P^{\downarrow}_{Y}(k)
  \quad\text{for all}\quad l=1,2,\dots,n-1.
\end{equation}
here $P^{\downarrow}_{X}$ means a reordering of the probability vector such that
\(P_{X}(k)\geq P_{X}(l)\) for$l>k$ and all $k$ and $l$. We then say that $P_{X}$
is majorized by $P_{Y}$.~\cite{Jorswieck2007,Marshall2011}
\begin{definition}
  A function $\phi:\mathbb{R}^n\to\mathbb{R}$ is said to be Schur-convex if
  \(x\prec y\) implies \(\phi(x)\leq\phi(y)\). Similarly a function is said to be
  Schur-concave if \(x\prec y\) implies \(\phi(y)\leq\phi(x)\).
\end{definition}

With this we state the following lemma which will be needed in the proofs of the
main results see~\cite{Marshall2011}.
\begin{lemma}%
  Let $P_{X}$ and $P_{Y}$ be probability distributions over the alphabet
  $\mathcal{X}=\mathcal{Y}$.  Assume that $P_{X}\prec P_{Y}$ then we have
  $H(X)\geq H(Y)$.
\end{lemma}

In the following we give an overview of the fundamental well known results of
rate distortion theory which will be needed in the main part of this paper.
These results can be found in the standard literature such
as~\cite{Cover2006,Yeung2008,Csiszar2011}. A more in depth treatment is found
in~\cite{Gallager1968} and~\cite{Berger1971}.

Let \({(X_{n})}_{n\in\mathbb{N}}\) be a discrete memoryless source over a
discrete and finite alphabet \(\mathcal{X}\). Another finite set \(\mathcal{Y}\)
will be used as the reproduction alphabet.

The sequences $y^{(n)}$ from the reproduction alphabet \(\mathcal{Y}^{n}\) are
used to code the sequences $x^{(n)}$ from the source alphabet
$\mathcal{X}^n$. The goal in rate distortion theory is to code the source by a
rate a small as possible while at the same time limiting the inevitable
distortion, defined by a distortion measure given between the sequences $x^{(n)}$
and $y^{(n)}$, to a certain level.

We define a single letter distortion measure by
\begin{equation}
  d:\mathcal{X}\times\mathcal{Y}\to\overline{\mathbb{R}}_{\geq0}.
\end{equation}
The average distortion between two sequences $x^{(n)}$ and $y^{(n)}$ is then
simply given by
\begin{equation}
  \bar{d}(x^{(n)},y^{(n)})=\frac{1}{n}\sum_{i=1}^{n}d(x_{i}, y_{i}).
\end{equation}
In this paper to simplify the treatment we will only consider finite distortion
measures.

A single letter distortion measure is said to be normal if for every
$x\in\mathcal{X}$ there exist an $y$ such that \(d(x,y)=0\).

To state the main results of rate distortion theory we further need the notion
of a source code.  A \((M_{n},f,g)\) source code of blocklength $n$ for the
discrete memoryless source \((X_{n})_{n\in\mathbb{N}}\) consist of an encoder
\begin{equation}
  f: \mathcal{X}^{n} \to\{1,\dots, M_{n}\}
\end{equation}
and  decoder
\begin{equation}
  g: \{1,\dots,M_{n}\}\to\mathcal{Y}^n
\end{equation}
such that after encoding and decoding, the source sequence $x^{(n)}$ is
reproduced as $y^{(n)}=f(g(x^{(n)}))$. The reproduction sequences are also
called codewords the set of all reproduction sequences for a given code is
called a codebook.

We define the $\epsilon$-fidelity criterion for a \((M_{n},f,g)\) source code if
\begin{equation}
  \Pr\{\bar{d}(X^{(n)},Y^{(n)})\leq D\}\geq 1+\epsilon
\end{equation}
holds. Another possibility often found in the literature is to define the
average fidelity criterion for a codebook by
\begin{equation}
  \mathbb{E}[\bar{d}(X^{(n)},Y^{(n)})] \leq D.
\end{equation}
As it can be shown that both criteria lead to the same results, in the source
coding theorems considered in this paper, we will only use the
$\epsilon$-fidelity criterion in the following.
  
\begin{definition}
  Given a distortion level $D\in\overline{\mathbb{R}}_{\geq0}$, a non-negative
  number $R$ is called $\epsilon$-achievable for a source
  \({(X_{n})}_{n\in\mathbb{N}}\) with respect to the distortion measure $d(k,l)$
  if for every $\delta>0$ there exist a sequence of \((M_{n},f,g)\) rate
  distortion codes such that
  \begin{equation}
    \frac{1}{n}\log(M_{n})\leq R+\delta
  \end{equation} 
  and the $\epsilon$-fidelity criterion holds for this code.
  
  Furthermore a rate $R$ is achievable if it is $\epsilon$-achievable for every
  $\epsilon>0$.  The pair of an achievable rate and distortion level $D$ is
  called achievable rate distortion pair $(R,D)$. Moreover the rate distortion
  function $R(D)$ will be defined as the infimum over all achievable rates $R$
  given a fixed distortion level of $D$.
\end{definition}
This definition leads to the following fundamental rate distortion theorem
due to Shannon~\cite{Shannon1948}~(cf.~\cite{Berger1971,Gallager1968}).
\begin{theorem}
  Let ${(X_{n})}_{n\in\mathbb{N}}$ be an identically distributed memoryless
  source over the alphabet $\mathcal{X}$ generated by the distribution $P_{X}$.
  Then for every single letter distortion measure
  \(d:\mathcal{X}\times\mathcal{Y}\to\mathbb{R}_{\geq0}\) and distortion level
  $D\geq0$ the rate distortion function is given by
  \begin{equation}
    R(D)= \inf_{P_{Y|X}:\mathbb{E}[d(X,Y)]\leq D}I(X;Y).
  \end{equation}
  Where $I(X;Y)$ is the mutual information between the source and the
  reproduction alphabet, minimized over all transition probabilities $P_{Y|X}$
  and the expected distortion level is given by
  \begin{equation}
    \mathbb{E}[d(X,Y)]=
    \sum_{x\in\mathcal{X}}\sum_{y\in\mathcal{Y}}P_{X}(x)P_{Y|X}(y|x)d(x,y).
  \end{equation}
\end{theorem}

\begin{definition}
  Let \(y\in\mathcal{Y}\) minimize the expected distortion we then
    define the maximal distortion by
    \begin{equation}
      D_{\max}=\min_{y\in\mathcal{Y}}\mathbb{E}[d(X,y)].
    \end{equation}
    An easy consequence of this definition is that $R(D)=0$ for all $D\geq
    D_{\max}$.\cite{Gallager1968},\cite{Berger1971}.
\end{definition}

\section{Analytic Calculation of the Rate Distortion Function}
To analyze the rate distortion function regarding computability the basic steps
towards an analytical ansatz to the solution of the rate distortion function in
the finite discrete case, from the literature will be discussed. These provide
important insights about what is theoretically known about the rate distortion
function and its optimizing conditional probability matrix in general and are
the basis to understand the computability of the rate distortion function and
its optimizer.

A good starting point for analytical calculation is found in the Russian
paper~\cite{Erokhin1958} by Erokhin. Erokhins results are only valid for the
Hamming distortion measure but provide basic ideas used in the calculation of
the rate distortion function. More complete results and analytical calculations
are found in the work of Gallager~\cite{Gallager1968} and
Berger~\cite{Berger1971}.

In the rest of this paper we will consider only discrete source alphabets
$\mathcal{X}$ of size $|\mathcal{X}|=K$ as well as reproduction alphabets
$\mathcal{Y}$ of size $|\mathcal{Y}|=L$. The nature of those alphabets is
irrelevant for the analytical part. We therefore assume without loss of
generality that both Alphabets are numbered by
$\mathcal{X}=\{0,1,2,\dots,K-1\}$ and $\mathcal{Y}=\{0,1,2,\dots,L-1\}$.

We sometimes define the distortion measures
\(d:\mathcal{X}\times\mathcal{Y}\to\bar{\mathbb{R}}_{\geq0}\) by an $L\times K$
matrix
\begin{equation}
  d(k,l)={(d_{k,l})}_{0\leq k\leq K-1,0\leq l\leq L-1}= 
  \begin{bmatrix} 
    d_{0,0}  &  d_{0,1} & \cdots  & d_{0,L-1}\\
    d_{1,0}  &  d_{1,1} & \cdots  & d_{1,L-1} \\
    \vdots &  \vdots & \ddots & \vdots \\
    d_{K-1,0}  & d_{K-1,1}  &  \cdots  & d_{K-1,L-1}
  \end{bmatrix}.
\end{equation}

The problem of minimization of mutual information with respect to a distortion
level $D$ in this setting is a well posed convex optimization problem subject to
the constraints
\begin{equation}
  0\leq P_{Y|X}(l|k)\qquad\forall l,k
\end{equation}
and
\begin{equation}
  0\leq P_{Y}(l)\leq1\qquad\forall l
\end{equation}
as well as
\begin{equation}
  \mathbb{E}[d(X,Y)]=\sum_{x=0}^{K-1}\sum_{y=0}^{L-1}P_{X}(k)P_{Y|X}(y|x)d(k,l)=D.
\end{equation}

The typical way to solve this is to use Lagrange multipliers to incorporate the
additional boundary conditions into the optimization problem.

Applying those, the minimization becomes
\begin{equation}
  F(\mu^K, \lambda, P_{Y|X})=I(P_{X}; P_{Y|X})-\sum_{k=0}^{K-1}\mu_{k}
  \sum_{l=0}^{L-1}P_{Y|X}(l|k)-\lambda\sum_{k=0}^{K-1}\sum_{l=0}^{L-1}P_{X}(k)P_{Y|X}(l|k)d(k,l).
\end{equation}

This then finally leads, after some elementary manipulations, to the following equations.
\begin{equation}\label{eq:21}
  1=\sum_{k=0}^{K-1}\mu_{k}\exp\left(-\lambda d(k,l)\right)\quad l=0,1,\dots L-1
\end{equation}

\begin{equation}
  P_{X}(k) = \mu_{k}\sum_{l=0}^{L-1}P_{Y}(l)\exp\left(-\lambda d(k,l)\right)\quad k=0,1,\dots K-1.
\end{equation}

These systems of linear equations in $\mu_{k}$ and $P_{Y}$, might depending on
the alphabet sizes $K$ and $L$ and the complexity of the distortion measure be
solved by standard methods of linear algebra. Nevertheless these systems are
typically non-linear in $\lambda$ which makes a solution at least analytically
mostly impossible.

From a solution of these systems the optimal transition probability can be
calculated by
\begin{equation}
  P_{Y|X}(l|k)=\frac{P_{Y}(l)\mu_{k}\exp\left(-\lambda d(k,l)\right)}{P_{X}(k)}\qquad k=0,1,\dots,K-1\quad l=0,1,\dots,L-1.
\end{equation}

Nevertheless in the steps of the above solution procedure it is difficult to
also incorporate the boundary conditions
\begin{equation}
  \sum_{k=1}^{K}P_{Y|X}(l|k)=1\quad0\leq l\leq L-1
\end{equation}
into the function $F$. This lead to solutions where $P_{Y}(l)<0$ might occur in
the above solution when the solution lies on the boundary of the constraint
set.\cite{Gallager1968}

The optimization is therefore complicated by leaving only a try and error
approach (see \cite{Gallager1968, Berger1971}). The following theorem
due to Gallager~\cite{Gallager1968}, and Berger~\cite{Berger1971} gives a
partly solution for those inconvenient cases.
\begin{theorem}\label{thm:gallager}
  Let the optimization problem be given as above. Then the rate distortion
  function is given by
  \begin{equation}\label{eq:30}
    R(D) = H(P_{X})+\max_{\lambda, \mu^K}\left[\sum_{k=0}^{K-1}P_{X}(k)\log(\mu_{k})-\lambda D\right]
  \end{equation}
  with 
  \begin{equation}
    H(X)=H(P_{X})=-\sum_{k=0}^{K-1}P_{X}(k)\log(P_{X}(k)).
  \end{equation}
  
  Where the maximization is over all $\mu_{k}$ such that
  \begin{equation}
    \sum_{k=0}^{K-1}\mu_{k}\exp\left(-\lambda d(k,l)\right)\leq1
    \label{eq:8}
  \end{equation}
  and equality is achieved in equation~\eqref{eq:8} for all $l$ for which  $P_{Y}(l)>0$ holds.
\end{theorem}

This gives in theory a solution procedure for the rate distortion function. But
the try and error notion of the above solution procedure is still present in
checking equation~\eqref{eq:8} and $P_{Y}>0$ for every possible solution. This
and the fact that the equation systems might be under determined or over
determined limit the above procedure for automated solution algorithms
significantly. As will be shown in the main part of this paper this kind of
behavior leads to computability problems.

\section{Prerequisites from Computable Analysis}
In the following we introduce the basic notions of computability and computable
analysis as needed in the main part of this paper. Most of the information here
can be found in greater depth in~\cite{PourEl2016, KLeene1971, Manin2010}.

The fundamental concept of a digital computer is completely described by Turing
machines introduced by Alan Turing~\cite{Turing1937} to solve the decision
problem.

Given a tape with an input sequence a Touring machine produces an output on an
output tape in a finite amount of computation steps controlled by a finite state
machine. The complexity and memory needed for this computations are not limited
in this theoretical model.

\begin{figure}[ht]
  \centering
  \includegraphics[width=.7\textwidth]{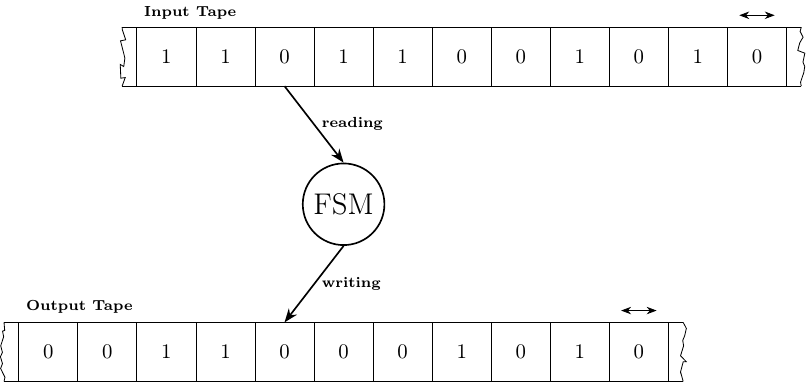}
  \caption{Turing machine~(cf.~\cite{Weihrauch2000})}\label{fig:turingm}
\end{figure}

\begin{figure}[ht]
  \centering
  \includegraphics[width=.6\textwidth]{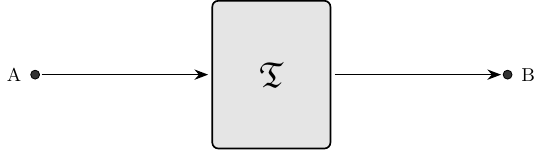}
  \caption{Block diagram of a Turing machine which gets a representation of $A$
    as input and computes a representation of $B$ as output.}\label{fig:turing-picto0}%
\end{figure}

Turing computable functions therefore describe all possible functions which can
be calculated by a modern digital computer using a discrete alphabet under no
time and memory constraints. One standard universal model for a Turing machine
is shown in figure~\ref{fig:turingm}. We will also use the simple block
diagram~\ref{fig:turing-picto0} to emphasize the computation of a Turing
machine.

 While this model gives a simple introduction into algorithmic computability
 here we also use the notion of recursive functions and sets as introduced
 in~\cite{Soare1987, Rogers1967, KLeene1971} for example.  These give a more
 mathematical way of describing the notion of a computability and are easier to
 handle in the proofs presented here. Moreover it has been shown the notion of
 recursive functions are identical to the notion of Turing computable functions
 and therefore describe in a mathematical notion all the function which can be
 computed on a digital computer~(cf.\cite{Manin2010}).

\begin{definition}
  A set $A\subset\mathbb{N}$ is called recursively enumerable if there exist a
  recursive function $g:\mathbb{N}\to\mathbb{N}$ such that for every \(m\in A\)
  there exist a $n\in\mathbb{N}$ such that $m=g(n)$.
\end{definition} 

\begin{definition}
  A set $A\subset\mathbb{N}$ is called decidable if its characteristic function
  $\chi_{A}$ is computable and therefore recursive.
\end{definition}

\begin{lemma}
  A set $A\subset\mathbb{N}$ is decidable if both $A$ and
  $A^{c}\subset\mathbb{N}$ are recursively enumerable.
\end{lemma}

\begin{theorem}
  There exist a set $A\subset\mathbb{N}$ which is not decidable or equivalently which is recursively
  enumerable but non-recursive.
\end{theorem}

In the following we extent the notion of computable functions from $\mathbb{N}$ to $\mathbb{Q}$ and
$\mathbb{R}$.

\begin{definition}{}
  A sequence of rational numbers \({(r_{k})}_{k\in\mathbb{N}}\) is called computable if
  there exist recursive functions
  \(s, a, b:\mathbb{N}\to\mathbb{N}\) such that $b(k)\neq0$ for
  $k\in\mathbb{N}$ and
  \begin{equation}
    r_{k}={(-1)}^{s(k)}\frac{a(k)}{b(k)} \quad\ k\in\mathbb{N}
  \end{equation}
\end{definition}
With this in mind we define the computable real numbers as
\begin{definition}
  A real number \(x\in\mathbb{R}\) is called computable if there exist a
  computable sequence of rational numbers ${(r_{k})}_{k\in\mathbb{N}}$ and a
  recursive function $e:\mathbb{N}\to\mathbb{N}$ such that
  \begin{equation}
    |x-r_{k}|\leq2^{-N}\quad\text{for all}\quad\ k\geq\ e(N)
  \end{equation}
\end{definition}
A sequence of rational numbers \({(r_{k})}_{k\in\mathbb{N}}\) where the modulus of
convergence can be controlled by a recursive functions $e(k)$ is called
effectively convergent. We denote the set of all computable real numbers by
$\mathbb{R}_{\mathrm{c}}$

This definition extends in the following obvious direction.
\begin{definition}
  A sequence of real numbers \({(x_{n})}_{{n}\in\mathbb{N}}\)
  \(x_{n}\in\mathbb{R}\) is called computable if there exist a computable double
  sequence of rational numbers ${(r_{n,k})}_{k\in\mathbb{N}}$ and a recursive
  function $e:\mathbb{N}\times\mathbb{N}\to\mathbb{N}$ such that
  \begin{equation}
    |x_{n}-r_{n,k}|\leq2^{-N}\quad\text{for all}\quad\ k\geq\ e(n,N)
  \end{equation}
\end{definition}
In the same notion we can now introduce the concept of computable real functions. 

\begin{definition}{Banach-Mazur Computable Function (see~\cite{PourEl2016, Weihrauch2000})}
  A real function \(f:\mathbb{R}_{\mathrm{c}}\to\mathbb{R}\) which maps every computable
  sequence of real numbers to a computable sequence of real numbers is called
  Banach-Mazur computable.
\end{definition}

As this notion of a computable function is typical too weak we additional
introduce the notion of a computable function in the sense of Richards and
Pour-El~\cite{PourEl2016}.
\begin{definition}
Let \(I^{n}\subset\mathbb{R}^{n}\) be the closed and bounded $n$-dimensional
rectangle defined by \(I^{n}=\{a_{i}\leq x_{i}\leq b_{i}, 1\leq i \leq n\}\)
with \(a_{i},b_{i}\in\mathbb{R}_{\mathrm{c}}\) computable real numbers for all
$i$.

Then a real function \(f:I^{n}\to\mathbb{R}\) is computable if
\begin{enumerate}
\item $f$ is Banach-Mazur computable and therefore maps every computable
  sequence of real numbers \((x_{k})_{k\in\mathbb{N}}\) with \(x_{k}\in I^{n}\)
  to a computable sequence of reals \((f(x_{k}))_{k\in\mathbb{N}}\)
\item There exist a recursive function \(d:\mathbb{N}\to\mathbb{N}\) such
  that for all \(x^{n},y^{n}\in I^{n}\) and all $N\in\mathbb{N}$ we have
  \begin{equation}
    {\|x^{(n)}-y^{(n)}\|}_{2}\leq \frac{1}{d(N)}\implies
 |f(x^{(n)})-f(y^{(n)})|\leq 2^{-N}.
  \end{equation}
  In this case $f$ is called \textit{uniformly effective continuous}.
\end{enumerate}
\end{definition}
This definition extends further in the obvious way to computable sequences of
functions. For a more detailed discussion see \cite{PourEl2016}. 
\begin{definition}
  Let \(I^{n}\subset\mathbb{R}^{n}\) be the closed and bounded $n$-dimensional
  rectangle defined by \(I^{n}=\{a_{i}\leq x_{i}\leq b_{i}, 1\leq i \leq n\}\)
  with \(a_{i},b_{i}\in\mathbb{R}_{\mathrm{c}}\) computable real numbers for all
  $i$.
  
  A sequence of functions \((f_{m})_{m\in\mathbb{N}}\) with
  \(f_{m}:I^{n}\to\mathbb{R}\) is computable if
  \begin{enumerate}
  \item for any computable sequence of points \((x_{k})_{k\in\mathbb{N}}\) in
    the compact rectangle \(x_{k}\in I^{n}\) the double sequence of reals given
    by \((f_{m}(x_{k}))_{k,n\in\mathbb{N}}\) is a computable sequence.
  \item there exist a recursive function \(d:\mathbb{N}\to\mathbb{N}\) such
    that for all \(x^{n},y^{n}\in I^{n}\) and all $m,N\in\mathbb{N}$ we have
    \begin{equation}
      {\|x^{(n)}-y^{(n)}\|}_{2}\leq \frac{1}{d(m,N)}\implies
      |f_{m}(x^{(n)})-f_{m}(y^{(n)})|\leq 2^{-N}.
  \end{equation}
  \end{enumerate}
\end{definition}
With this information we can now give the following important result
(see~\cite{PourEl2016, Specker1959}).

\begin{theorem}
  Let $(f_{m})_{m\in\mathbb{N}}$ with $f_{m}:I^{n}\to\mathbb{R}$ be a computable
  sequence of functions defined on the compact $n$-dimensional rectangle
  $I^{n}$. Then the sequence of minima 
  \begin{equation}
    f_{m,\min}=\min_{x^{(n)}\in I^{n}}f_{m}(x^{(n)})
  \end{equation}
  on the compact set $I^{n}$ is a computable sequence of real numbers.
\end{theorem}
Nevertheless while the sequence of minima in this case are computable the
sequences of points $x^{(n)}_{m,\min}$ where the minima are attained need not be
computable not even for a single function (see~\cite{Specker1959,PourEl2016}).

\begin{remark}
  Computability by some general Turing machine is shown is shown in
  figure~\ref{fig:turing-picto0}. This Turing machine is getting a
  representation of computable numbers, probability densities or distortions as
  input and computes a representation of a solution as the result. The
  representations in this case must be in the form of a finite string of input
  symbols from an input alphabet. This process is shown with more details in
  figure~\ref{fig:turingm}.

  In our case we are interested in the computation of a representation of an
  optimal test channel given a representation of a source distribution,
  distortion measure and distortion level.
\end{remark}

\section{Main Results and Discussion}
In this section we give and discuss the main results of our findings. We will
give results regarding the computability of the optimizer for the rate
distortion function $R(D)$ for arbitrary computable input distributions $P_{X}$
and distortion measures $d$ on the finite source and reconstruction alphabets
$\mathcal{X}$ and $\mathcal{Y}$. The set of all computable source distributions
on the input alphabet $\mathcal{X}$ is denoted by
$\mathcal{P_{\mathrm{c}}}(\mathcal{X})$. We will only treat finite distortion
measures to simplify the treatment. Most of the results hold in a similar
fashion also for distortion measures which allow infinite distortions for some
input and reconstruction letters.

We start by giving the following simple finding for general computable distortion
measures.
\begin{lemma}
  Let \(d:\mathcal{X}\times\mathcal{Y}\to\mathbb{R}_{\geq0,\mathrm{c}}\) be a
  computable single letter distortion measure on the arbitrary but finite
  alphabets $\mathcal{X}$ and $\mathcal{Y}$. Moreover assume that the source
  distribution is computable \(P_{X}\in\mathcal{P}_{c}(\mathcal{X})\).
  
  Then there exist a normal, computable, single letter distortion measure $d'$
  having the same solution and optimizer as $d$ except that its rate distortion
  function is shifted by a computable number $D'\in\mathbb{R}_{\mathrm{c}}$.
  \begin{proof}
    Because $d$ is a computable function and $\mathcal{Y}$ is finite the minimum
    \begin{equation}
      c_{k}=\min_{l\in\mathcal{Y}} d(k,l)
    \end{equation}
    of $d$ is computable for every fixed $k\in\mathcal{X}$. With that we define
    \begin{equation}
      d':=d(k,l)-c_{k}
    \end{equation}
    as another distortion measure. It follows that $d'$ is a normal distortion
    measure. Computability follows from the computability of $c_{k}$ and $d$
    together with the assumption of a finite source alphabet $\mathcal{X}$.  Now
    assume that the rate distortion function of $d$ is given by $R(D)$. Then the
    same transition probability will also minimize $R'$ for $d'$ but with
    shifted distortion level given by
    \begin{align}
      D'=&\mathbb{E}[d'(X,Y)]=\sum_{k\in\mathcal{X}}\sum_{l\in\mathcal{Y}}p(k,l)(d(k,l)-c_{k})\\
      =&D-\mathbb{E}_{X}[c_{k}].
    \end{align}
    This follows from standard results in rate distortion
    theory~\cite{Gallager1968, Berger1971}. We can therefore always assume that
    the considered distortion measures are normal.
  \end{proof}
\end{lemma}

As is easily proofed mutual information is a continuous function when the input
$\mathcal{X}$ and reconstruction alphabet $\mathcal{Y}$ are
finite.\cite{Yeung2008} Moreover mutual information is a computable function
given a computable input probability vector and computable transition
probability matrix. From this and the fact that mutual information is defined on
a compact set we see that also $R(D)$ is a computable function as it is the
minimum of a computable function. Another insightful way to see this is to
examine the solution ansatz and especially equation~\eqref{eq:30}. Nevertheles
note that this does not necessarily hold for an optimizing test channel
probability $P^{*}_{Y|X}$.

\begin{figure}[ht]
  \centering
  \includegraphics[width=.6\textwidth]{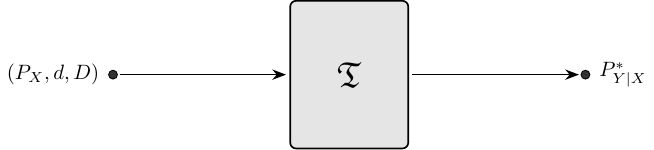}
  \caption{Block diagram of a Turing machine computing $P^{*}_{Y|X}$. The Turing
    machine gets a representation for $P_{X}$, $d$ and $D$ as input and computes
    a representation of $P^{*}_{Y|X}$ as output.}\label{fig:turing-picto1}
\end{figure}

We are therefore interested in the construction of a function
$F_{\mathrm{opt}}$ which for a given a computable distortion measure
$d:\mathcal{X}\times\mathcal{Y}\to\mathbb{R}_{\mathrm{c},\geq0}$, distortion
level $D$ and computable source distribution
\(P_{X}\in\mathcal{P}_{\mathrm{c}}(\mathcal{X})\) computes an optimizing
transition probability $P^{*}_{Y|X}$. This is equivalent to the Turing
machine depicted in the block diagram in figure~\ref{fig:turing-picto1}.

In general there might be more than one $P^{*}_{Y|X}$. This is for example
always the case when the the rate distortion function is not a strictly convex
function of the transition probability. For this we denote all computable
optimal transition probabilities for a set of $d(k,l)$, $D$ and source
probability vector $P_{X}$ by
\(\mathcal{P}_{\mathrm{c},\mathrm{opt}}(d,D,P_{X})\). We are then interested in
the computability of a function
\(F_{\mathrm{opt}}:(P_{X},d,D)\to\mathcal{P}_{\mathrm{opt}}\) which computes an
optimal transition probability
\(P^{*}_{Y|X}\in\mathcal{P}_{\mathrm{c},\mathrm{opt}}(d,D,P_{X})\) given an
arbitrary computable source distribution
$P_{X}\in\mathcal{P}_{\mathrm{c}}(\mathcal{X})$, computable distortion measure
$d:\mathcal{X}\times\mathcal{Y}\to\mathbb{R}_{\geq0,\mathrm{c}}$ and computable
distortion level $D\in\mathbb{R}_{\mathrm{c}}$. Let
\(\mathcal{M}_{\mathrm{opt}}(\mathcal{X}, \mathcal{Y})\) be the set of all such
functions $F_{\mathrm{opt}}$.We are then interested in the computability of any
of those functions.

With this we give the main result of this paper.
\begin{corollary}
  Let $\mathcal{M}(\mathcal{X},\mathcal{Y})$ be the set of all functions
  \(F_{\mathrm{opt}}:(P_{X},d,D)\to\mathcal{P}_{\mathrm{opt}}\) getting a
  computable source distribution
  $P_{X}\in\mathcal{P}_{\mathrm{c}}(\mathcal{X})$, distortion measure
  \(d:\mathcal{X}\times\mathcal{Y}\to\mathbb{R}_{\geq0, \mathrm{c}}\) and
  computable distortion level $D\in\mathbb{R}_{\mathrm{c}}$ as input and
  generating any \(P^{*}_{Y|X}\in\mathcal{P}_{\mathrm{opt}}(d,D,P_{X})\) as
  output. Then for arbitrary but finite $\mathcal{X},\mathcal{Y}$ with
  $|\mathcal{X}|\geq2$ and $|\mathcal{Y}|\geq2$ and if the source probability
  $P_{X}$ is non trivial there exist no function
  \(F_{\mathrm{opt}}\in\mathcal{M}(\mathcal{X},\mathcal{Y})\) that is
  Banach-Mazur and therefore Turing computable.
\end{corollary}

This corollary shows that the approach of finding a general closed form solution
or even some algorithm for computing the optimizer for the rate distortion
function is not possible at all. The proof of this theorem will follow as a
simple corollary from the following more specific theorems.

Any of the functions $F_{\mathrm{opt}}$ as defined above would if computable
provide an algorithm to compute the optimizing $P^{*}_{Y|X}$ in the most general
case. That means it would give an algorithm for a Turing machine to compute the
optimizer given arbitrary but computable input distribution, distortion measure
and distortion level as input.

\begin{figure}[ht]
  \centering
  \includegraphics[width=.6\textwidth]{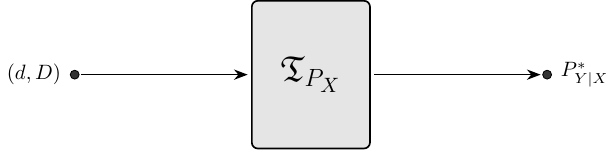}
  \caption{Block diagram of a Turing machine computing a representation of
    $P^{*}_{Y|X}$ as output. Now the Turing machine only gets a representation
    of $d$ and $D$ as Input. Even if there exist a Turing machine of this type
    for every $P_{X}$ there does not necessarily exist a Turing machine of the
    more general kind shown
    in figure~\ref{fig:turing-picto1}.}\label{fig:turing-picto2}
\end{figure}

In the following we start by analyze the computability of any of the
$F_{\mathrm{opt}}$ under more relaxed and specific conditions. For this we
first fix the source probability distribution to be any
$P_{X}\in\mathcal{P}_{\mathrm{c}}(\mathcal{X})$ with support on the set
$\mathcal{X}$. We are therefore at first limiting our discussion to the subset
\(\mathcal{M}(\mathcal{X},\mathcal{Y}, P_{X})\subset\mathcal{M}(\mathcal{X},
\mathcal{Y})\). This has the important implication that a function
$F_{\mathrm{opt}}$ in this set would allow us to calculate a $P^{*}_{Y|X}$ only
for one fixed $P_{X}\in\mathcal{P}_{\mathrm{c}}(\mathcal{X})$ while another input
distribution might require another function or algorithm. A Turing
machine for this case is shown in figure~\ref{fig:turing-picto2}

This means we want to analyze the computability of $F_{\mathrm{opt}}$ for
changing distortion measures $d$. Here we are searching for a function which
takes a distortion measure
\(d:\mathcal{X}\times\mathcal{Y}\to\mathbb{R}_{\mathrm{c},\geq0}\) and
computable distortion level $D\in\mathbb{R}_{\mathrm{c}}$ as argument while
considering the computable source probability
$P_{X}\in\mathcal{P}_{\mathrm{c}}(\mathcal{X})$ and alphabets $\mathcal{X}$,
$\mathcal{Y}$ as fixed. The following theorem gives a definite answer to the
computability of such a function when only the distortion measure is variable
input to such a function.

\begin{theorem}\label{thm:main-d}
  Let a finite source alphabet $\mathcal{X}$ with \(|\mathcal{X}|=K\geq2\) and a
  finite reproduction alphabet $\mathcal{Y}$ with \(|\mathcal{Y}|=K+2\) be
  given. Then for every computable probability distribution on the source
  alphabet $P_{X}\in\mathcal{P}_{\mathrm{c}}(\mathcal{X})$ with support on
  $\mathcal{X}$ there is no function
  \(F_{\mathrm{opt}}\in\mathcal{M}(\mathcal{X},\mathcal{Y},P_{X})\) that is
  Banach-Mazur and therefore Turing computable.
\end{theorem}

We begin the proof by analyzing the behavior for any of the functions
\(F_{\mathrm{opt}}\in\mathcal{M}(\mathcal{X},\mathcal{Y},P_{X})\). The theorem
then follows from the following Lemma.
\begin{lemma}
  Let a finite source alphabet $\mathcal{X}$ with \(|\mathcal{X}|=K\geq2\) and a
  finite reproduction alphabet $\mathcal{Y}$ with \(|\mathcal{Y}|=K+2\) be given.
  
  Then for every computable probability distribution on the source alphabet
  $P_{X}\in\mathcal{P}_{\mathrm{c}}(\mathcal{X})$ with \(0<P_{X}(k)<1\) for
  every \(0\leq k \leq K-1\) there exist a computable sequence of normal, single
  letter distortion measures
  \(d_{n}:\mathcal{X}\times\mathcal{Y}\to\mathbb{R}_{\mathrm{c},\geq0}\) and a
  computable minimal distortion level $D_{\min}\in\mathbb{R}_{\mathrm{c},\geq0}$
  such that the sequence of optimal transition probabilities
  $P^{*}_{Y|X,n}\in\mathcal{P}_{\mathrm{opt}}(d_{n},D,P_{X})$ given by any of
  the $F_{\mathrm{opt}}(d_{n}, D, P_{X})$ is not Banach-Mazur computable for
  every computable distortion level between \(D_{\min}\leq D\leq D_{\max}\).
  \begin{proof}
    \begin{figure}[!ht]
      \centering
      \includegraphics[width=.3\textwidth]{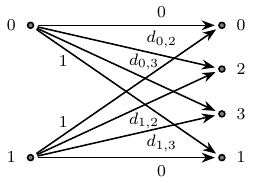}
      \caption{Erasure distortion measure with a second erasure symbol and $|\mathcal{X}|=2$, $|\mathcal{Y}|=4$}\label{img:erasure-distortion}
    \end{figure}

    We start the proof by the construction of an appropriate, computable
    distortion measure \(d:\mathbb{N}\times\mathbb{N}\to\mathbb{R}_{\geq
      0,\mathrm{c}}\).  This measure is the multidimensional analogue to the
    erasure distortion measure as analyzed for example by
    Berger~\cite{Berger1971}.
    \begin{equation}
      d(k,l)=
      \begin{cases}
        0,& \text{if}\quad k=l\\
        d_{1},& \text{if}\quad k\neq l,\; l=K\\
        d_{2},& \text{if}\quad k\neq l,\; l=K+1\\
        1, & \text{otherwise.}
      \end{cases}
    \end{equation}
    Another difference is the second additional erasure symbol. Both erasure
    symbols $K$ and $K+1$ have two different erasure symbol distortions $d_{1}$
    and $d_{2}$ while the distortion measure for the other source letters is
    given by the Hamming distortion measure.
    
    An image of this distortion measure in the case of $\mathcal{X}=2$ and
    $\mathcal{Y}=4$ is depicted in figure~\ref{img:erasure-distortion}.
    
    To show the non-computability we will construct sequences of this distortion
    measure by changing the erasure distortions $d_{1}$ and $d_{2}$ accordingly.
    
    We begin the proof by first giving an analytical solution ansatz for the
    section of the $R(D)$ for which we have $P_Y(k)>0$ for all $0\leq k\leq
    K-1$. This corresponds to the first section of the curve as can be easily
    seen by comparing the transition probabilities for $D=0$ and $D_{\max}$.
    
    We start by analyzing the solution ansatz as given by
    Theorem~\ref{thm:gallager}. The Lagrange equations as given by~\eqref{eq:21}
    then become
    \begin{align}
      \mu_{l}+\sum_{\substack{k=0\\k\neq l}}^{K-1}\mu_{k}e^{-\lambda}&\leq1\qquad 0\leq l\leq K-1\\
      e^{-d_{1}\lambda}\sum_{k=0}^{K-1}\mu_{k}&\leq1\qquad l=K\label{eq:10}\\
      e^{-d_{2}\lambda}\sum_{k=0}^{K-1}\mu_{k}&\leq1\qquad l=K+1\label{eq:11}.
    \end{align}
    Notice that only one of equation~\eqref{eq:10} or~\eqref{eq:11} can hold
    simultaneously with equality if $d_{1}\neq d_{2}$. So it is enough to
    analyze the \(R(D)\)-function for the case $P_{Y}(K+1)=0$. The other case
    will then follow accordingly. The important case $d_{1}=d_{2}$ will be
    analyzed in the end.
    
    We proceed by analyzing the first part of the $R(D)$ curve. We will show
    that the erasure symbol probability on the reproduction alphabet
    $P_{Y}(K)>0$ will be met in this first part if the erasure symbol distortion
    $d$ is choosen appropriately. We will then choose $D_{\min}$ to be any point
    on the abcissa of the first part of the $R(D)$ curve for which we have
    $P_{Y}(K)>c$ where $a$ is an arbitrary lower bound.
    
    For this we start by setting the erasure symbol distortion
    \(d_{1}\in\mathbb{R}_{\mathrm{c},\geq0}\) to be in the range $0\leq
    d_{1}<\min_{k}P_{X}(k)$. Then the equations for the reproduction probabilies
    become
    \begin{equation}
      \frac{P_{X}(k)}{\mu_{k}}=P_{Y}(k)+e^{-\lambda}(1-P_{Y}(k)-P_{Y}(K))+e^{-\lambda d_{1}}P_{Y}(K)
      \label{eq:17}\qquad 0\leq k\leq K-1.
    \end{equation}
    here we also used that the output probabilities must add up to $1$.
    
    Starting at $D=0$ we find that equation~\eqref{eq:10} is fulfilled with
    equality for all $l$ we therefore have $\mu_{k}=\mu_{0}$. Solving these
    equations we get
    \begin{equation}
      \mu_{0}=\mu_{k}=\frac{1}{1+(K-1)e^{-\lambda}}\label{eq:13}.
    \end{equation}
    
    For these low distortions $D$ near zero, equation~\eqref{eq:11} is not yet
    achieved with equality and we therefore have $P_{Y}(K)=0$. We then can
    simplify~\eqref{eq:17}.
    \begin{equation}
      P_{Y}(k)=\frac{P_{X}(k)(1+(K-1)e^{-\lambda})-e^{-\lambda}}{1-e^{-\lambda}}
    \end{equation}
    
    From this we also find the condition for which $P_{Y}(k)\geq0$ for $0\leq
    k\leq K-1$ given by
    \begin{equation}
      P_{Y}(k)\geq0 \iff P_{X}(k)\geq \frac{e^{-\lambda}}{1+(K-1)e^{-\lambda}}
    \end{equation}
    
    The distortion for these probabilities is found to be 
    \begin{equation}
      D=\frac{e^{-\lambda}}{1+(K-1)e^{-\lambda}}
    \end{equation}
    solving this for $\lambda$ we get
    \begin{equation}
      \lambda=\ln(K-1)+\ln\left(\frac{1-D}{D}\right).
    \end{equation}
    This result is identically to the well known Hamming distortion case
    (see~\cite{Gallager1968}).
    
    Now because $\lambda$ is decreasing with $D$ increasing there is a point
    $\lambda^*$ such that also equation~\eqref{eq:11} is fulfilled with equality
    \begin{equation}
      1=e^{-d_{1}\lambda}K\mu_{0}\label{eq:12}
    \end{equation}
    and we have $P_Y{K}\geq0$.  Now by using $\mu_{0}$ from
    equation~\eqref{eq:13} in equation~\eqref{eq:12} we get an estimate for
    $\lambda^{*}$.
    \begin{equation}
      f(\lambda^{*})=1+(K-1)e^{-\lambda^*}-Ke^{-d_{1}\lambda^*}=0\label{eq:14}
    \end{equation}
    Unfortunately this equation is transcendental and does not have an analytic
    solution. Nevertheless from
    \begin{equation}
      (K-1)e^{-\lambda^*}\geq0
    \end{equation}
    it is easy to see that
    \begin{equation}
      1-Ke^{-\lambda^* d_{1}}\leq0
    \end{equation}
    and therefore we have the following easy upper bound
    \begin{equation}
      \lambda^*\leq\frac{1}{d_{1}}\log(K).
    \end{equation}
    By differentiation of $f$ we further have
    \begin{equation}
      \frac{\mathrm{d}f(\lambda)}{\mathrm{d}\lambda}=(1-K)e^{-\lambda}-Kd_{1}e^{-d_{1}\lambda}
    \end{equation}
    as well as
    \begin{equation}
      \frac{\mathrm{d}^2f(\lambda)}{\mathrm{d}\lambda^2}=(K-1)e^{-\lambda}-Kd_{1}^{2}e^{-d_{1}\lambda}
    \end{equation}
    It therefore follows from $d_{1}\leq \frac{1}{K}$ that $f$ is zero for
    $\lambda=0$ then decreases until reaching a negative minimum at
    \begin{equation}
      \lambda=\frac{1}{1-d_{1}}\log\left(\frac{K-1}{d^2K}\right).
    \end{equation}
    Because this is the only extremum the curve is then only increasing crossing
    zero for $\lambda^*$.
    
    Giving the following possible estimate for $\lambda^*$
    \begin{equation}
      \frac{1}{1-d_{1}}\log\left(\frac{K-1}{d_{1}^2K}\right)\leq\lambda^*\leq\frac{1}{d_{1}}\log(K).
      \label{eq:23}
    \end{equation}
    It is seen by the intermediate value theorem that $\lambda^*$ is a
    computable real number for $d_{1}$.
    
    We further see that $\mu_{k}$ is still given by equation~\eqref{eq:13}. This
    allows us to compute the $P_{Y}$.
    \begin{equation}
      P_{Y}(k)=\frac{P_{X}(k)(1+(K-1)e^{-\lambda})-e^{-\lambda}-(e^{-\lambda d_{1}}-e^{-\lambda})P_{Y}(K)}{1-e^{-\lambda}}
    \end{equation}
    
    We further need to solve the equation for $D$ to get a solution for
    $P_{Y}(K)$ as this system is under determined and still dependent on
    $P_{Y}(K)$.
    \begin{align}
      D=&d_{1}\sum_{k=0}^{K-1}\mu_{k}e^{-\lambda d_{1}}P_{Y}(K)+\sum_{l=0}^{K-1}\sum_{\substack{k=0\\k\neq l}}^{K-1}\mu_{k}e^{-\lambda}P_{Y}(l)\\
      =&d_{1}K\mu_{0}P_{Y}(K)e^{-\lambda d_{1}}+e^{-\lambda}(K-1)\mu_{0}(1-P_{Y}(K))
    \end{align}
    
    Now by further using the equation for $\mu_{k}$ as well as the
    equation~\eqref{eq:14} for $\lambda^{*}$ we find the solution for
    $P_{Y}(K)$.
    \begin{equation}
      P_{Y}(K)=\frac{D[1+(K-1)e^{-\lambda^*}]-(K-1)e^{-\lambda^*}}{d[1+(K-1)e^{-\lambda^*}]-(K-1)e^{-\lambda^*}}\label{eq:20}
    \end{equation}

    We now finally want to select a $D_{\min}$ in such a way that $P_{Y}(K)>0$
    for every $D\geq D_{\min}$. From~\eqref{eq:20} note that $P_{Y}(K)>0$ if
    \begin{equation}
      D>\frac{(K-1)e^{-\lambda^*}}{1+(K-1)e^{-\lambda^*}}
    \end{equation}
    because $D\leq d_{1}$. To show that the solution we just attained is correct
    we further need to show that we still have $P_{Y}(k)\geq0$ for every $0\leq
    k\leq K-1$ as otherwise the equations and $\mu_{k}$ by
    Theorem~\ref{thm:gallager} do not hold with equality anymore.
    
    \begin{equation}
      P_{X}(k)\geq\frac{e^{-\lambda^*}}{1+(K-1)e^{-\lambda^*}}+
      \frac{e^{-\lambda^*d_{1}}-e^{-\lambda^*}}{1+(K-1)e^{-\lambda^*}}P_{Y}(K)
    \end{equation}
    It is clear that because $\lambda^*$ is fixed and positive that $P_{Y}(K)>0$ if 
    \begin{equation}
      \min_{k}P_{X}(k)>\frac{e^{-\lambda^*}}{1+(K-1)e^{-\lambda^*}}.
    \end{equation}
    Using the estimate for $\lambda^*$ from equation~\eqref{eq:23} and solving
    for $d_{1}$ we see that this is the case as long as
    \begin{equation}
      \frac{\log(K)}{\log\left({\left[\min_{k}P_{X}(k)\right]}^{-1}-K+1\right)}>d_{1}
    \end{equation}
    and therefore always the case because \(d_{1}<\min_{k}P_{X}(k)\leq\frac{1}{K}\).
    
    We now change the above derivation for the case $d_{1}=d_{2}$. Carefully
    examine the argumentation above we find
    \begin{equation}
      P_{Y}(K)+P_{Y}(K+1)=
      \frac{D[1+(K-1)e^{-\lambda^*}]-(K-1)e^{-\lambda^*}}{d[1+(K-1)e^{-\lambda^*}]-(K-1)e^{-\lambda^*}}
      \label{eq:25}
    \end{equation}
    
    \begin{equation}
      D=d_{1}K\mu_{0}\left[P_{Y}(K)+P_{Y}(K+1)\right]e^{-\lambda d_{1}}+e^{-\lambda}(K-1)\mu_{0}(1-P_{Y}(K)-P_{Y}(K+1)).
    \end{equation}
    
    Because the system of equations is under determined and therefore can only
    solved uniquely for $P_{Y}(K)+P_{Y}(K+1)$. Moreover it is seen by comparing
    the solutions that the solution for $P_{Y}(K)$ and $d_{2}<d_{1}$ can be
    obtained by interchanging the role with $P_{Y}(K+1)$ in the analysis. In the
    end the solution for $d_{1}=d_{2}$ is given by the convex combinations of
    the single solutions.
    
    Finally we will need the following simple lemma which is interesting on its
    own.
    \begin{lemma}\label{lem:01}
      Let a channel with one erasure symbol $K$ as above be given. Moreover let
      $D^*$ be any minimal distortion for which $P_{Y}(K)>0$ where $P_{Y}$ is
      the optimal reproduction probability vector. Then $P_{Y}(K)$ is monotone
      increasing for all $D\geq D^*$.
    \end{lemma}
    \begin{proof}
      We start by rewriting the optimization equation of the rate distortion
      problem.
      \begin{equation}
        R(D)= \inf_{P_{Y|X}:\mathbb{E}[d(X,Y)]\leq D}I(X;Y)=H(X)-\max_{P_{Y|X}:\mathbb{E}[d(X,Y)]\leq D}H(X|Y)
      \end{equation}
      Breaking the maximization further down we have
      \begin{equation}
        \max_{P_{Y|X}:\mathbb{E}[d(X,Y)]\leq D}H(X|Y)= \max_{P_{Y|X}:\mathbb{E}[d(X,Y)]\leq D}\sum_{y\in\mathcal{Y}}P_{Y}(y)H(X|Y=y).
      \end{equation}
      Now because $P_{Y}(K)>0$ we know that for $y=K$ the optimal conditional
      probability $P_{X|Y}$ is given by
      \begin{equation}
        P_{X|Y}(x|K)=\frac{1}{|\mathcal{X}|}
      \end{equation}
      This is because entropy and therefore also conditional entropy is
      maximized by a uniform distribution.\cite{Yeung2008} Because the erasure
      symbol has the same symbol distortion for every letter of $\mathcal{X}$
      using any other conditional distribution for $P_{X|Y}(k|K)$ would lead to
      an increase $R(D)$ but not $D$.
      
      Now assume that $D>D^*$ is further increased. We want to show that a
      decrease in $P_{Y}(K)$ leads to a contradiction in this case. But this
      follows directly because we know that the distribution $P_{X|Y}$
      conditional on $Y=K$ is given by a uniform distribution and we therefore
      find that
      \begin{equation}
        P_{X|Y}(\cdot|K)\prec P_{X|Y}(\cdot|k)\quad\text{for}\quad0\leq k\leq K-1
      \end{equation}
      and therefore by Lemma~\ref{lem:01}
      \begin{equation}
        H(X|Y=k)\leq H(X|Y=K)\quad\text{for}\quad0\leq k\leq K-1.
      \end{equation}
      From this we see that increasing $P_{Y}(K)$ leads to a smaller mutual
      information and therefore a smaller rate contradicting the assumption that
      decreasing $P_{Y}(K)$ is optimal.  This shows that $P_{Y}(K)$ is monotone
      increasing for all $D>D*$.
    \end{proof}
    
    We now are in a position to analyze the computability of the optimizing
    transition probabilities.
    
    This is done by the construction of two sequences of distortion
    measures. For this we choose any computable \(d\in\mathbb{R}_{\mathrm{c}}\)
    such that $d\leq\min_{k}P_{X}(k)$ holds.  We then take any recursively
    enumerable non-recursive set $\mathcal{A}$. For that there exist a recursive
    function $a(m)$ with \(\mathrm{range}(a)=\mathcal{A}\). We then define
    the sequence 
    \begin{equation}\label{eq:xnm}
      x_{n,m}:= 
      \begin{cases}
        2^{-i},& \text{if}\quad n=a(i),\text{ for } 0<i\leq m\\ 
        0&\text{otherwise.}
      \end{cases}
    \end{equation}

    It is easy to see that $2^{i}$ can be calculated from the knowledge of
    $a(i)$ by recursion and the sequence $x_{n,m}$ is therefore a computable
    double sequence of rationals.

   We continue by defining two sequences of computable distortion measures by
    \begin{equation}
      d^{(1)}_{n,m}(k,l)=
      \begin{cases}
        0,& \text{if}\quad k=l\\
        d,& \text{if}\quad k\neq l,\; l=K\\
        d+x_{n,m},& \text{if}\quad k\neq l,\; l=K+1\\
        1, & \text{otherwise.}
      \end{cases}
    \end{equation}
    and
    \begin{equation}
      d^{(2)}_{n,m}(k,l)=
      \begin{cases}
        0,& \text{if}\quad k=l\\
        d+x_{n,m},& \text{if}\quad k\neq l,\; l=K\\
        d,& \text{if}\quad k\neq l,\; l=K+1\\
        1, & \text{otherwise.}
      \end{cases}
    \end{equation}
    It is clear that all entries in \(d^{(1)}_{n,m}\), and \(d^{(2)}_{n,m}\) are
    computable sequences of rationals. To show that the sequences are computable
    sequences of computable matrices we need to show that the convergence
    \begin{equation}
      \lim_{m\to\infty}d^{(2)}_{n,m}=d^{(1)}_{n}
    \end{equation}
    
    \begin{equation}
      \lim_{m\to\infty}d^{(2)}_{n,m}=d^{(2)}_{n}
    \end{equation}
    is effective in $m$.
    
    For this fist consider that $i\leq m$ From this we simply get
    \begin{equation}
      \left\|d^{(1)}_{n,m}-d^{(1)}_{n}\right\|_{2}=\left\|d^{(2)}_{n,m}-d^{(2)}_{n}\right\|_{2}=0
    \end{equation}
    from the definition of $x_{n,m}$.
    
    Now observe that in the approximation of $x_{n}$ by $x_{n,m}$, an error is
    made only if we have $n=a(i)$ for some a $i>m$. In this case we get the
    upper bound $|x_{n}-x_{n,m}|\leq2^{-i}<2^{-m}$. It then follows that for
    $m=M+\frac{1}{2}\log(K)$ we have
    \begin{align}
      \left\|d^{(1)}_{n,m}-d^{(1)}_{n}\right\|_{2}=&\left\|d^{(2)}_{n,m}-d^{(2)}_{n}\right\|_{2}\\
      =&\sqrt{K{\left(x_{n,M+\frac{1}{2}\log(K)}-x_{n}\right)}^{2}}\\
      <&K^{\frac{1}{2}}2^{-M+\frac{1}{2}\log(K)}\\
      =&2^{-M}
    \end{align}
    which shows that the convergence of the distortion measures is effective in
    $m$ for every dimension of the input probability vector $K$.
    
    We want to show that $F_{\mathrm{opt}}$ cannot be a computable function for
    every $D\geq D_{\min}$. 
    
    For that assume that $F_{\mathrm{opt}}$ is Banach-Mazur computable.  We will
    show that this assumption leads to a contradiction. For this select a $D\geq
    D_{\min}$ this is possible because $D_{\min}$ is a computable number and hence
    can be approximated to an arbitrary precision. Moreover this $D$ can be selected
    in such a way that $P_{Y}(k)\geq0$ for every $0\leq k\leq K-1$ and such that
    $P_{Y}(K)$ or $P_{Y}(K+1)$ respectively are bounded from below by an arbitrary
    constant $c$. Again such a procedure is possible because the reproduction
    probability vector $P_{Y}$ for this path of the curve is computable by the above
    procedure and therefore can be approximated to an arbitrary precision.
    
    Now assume that $n\in\mathcal{A}$ so that $x_{n}=0$. We then have
    $d^{(1)}_{n}=d^{(2)}_{n}$ and the soltution for both distortion measures is
    given by all linear combinations of $P_{Y,n}(K)+P_{Y,n}(K+1)$ given
    by~\eqref{eq:25}. Denote a possible optimal transition probability given
    by $F_{\mathrm{opt}}$ in this case by $P^*_{X|Y,n}$. On the other hand if
    $n\notin\mathcal{A}$ we find that the solutions for $P^{(1)}_{Y,n}$ and
    $P^{(2)}_{Y,n}$ are different and this implies leads to
    \begin{align}
      2c&<\left|P^{(1)}_{Y,n}(K)-P^{(2)}_{Y,n}(K)\right|+\left|P^{(1)}_{Y,n}(K+1)-P^{(2)}_{Y,n}(K+1)\right|\\
      &=\left\|P^{(1)}_{Y,n}-P^{(2)}_{Y,n}\right\|_{\mathrm{TV}}\\
      &=\left|\sum_{y\in\mathcal{Y}}\sum_{x\in\mathcal{X}}P^{(1)}_{Y|X,n}(y|x)P_{X}(x)-P^{(2)}_{Y|X,n}(y|x)P_{X}(x)\right|\\
      &\leq\sum_{y\in\mathcal{Y}}\sum_{x\in\mathcal{X}}P_{X}(x)\left|P^{(1)}_{Y|X,n}(y|x)-P^{(2)}_{Y|X,n}(y|x)\right|\\
      &\leq\sum_{x\in\mathcal{X}}P_{X}(x)\max_{x\in\mathcal{X}}\sum_{y\in\mathcal{Y}}\left|P^{(1)}_{Y|X,n}(y|x)-P^{(2)}_{Y|X,n}(y|x)\right|\\
      &\leq\max_{x\in\mathcal{X}}\sum_{y\in\mathcal{Y}}\left|P^{(1)}_{Y|X,n}(y|x)-P^{(2)}_{Y|X,n}(y|x)\right|\\
      &=\left\|P^{(1)}_{Y|X,n}-P^{(2)}_{Y|X,n}\right\|_{\mathrm{TV}}.
    \end{align}
    Because all the involved functions are computable and because
    $c\in\mathbb{Q}$ we get the following bound for arbitrary $n$.
    
    \begin{align}
      2c&<\left\|P^{(1)}_{Y|X,n}-P^{(2)}_{Y|X,n}\right\|_{TV}\\
      &=\left\|P^{(1)}_{Y|X,n}+P^{*}_{Y|X,n}-P^{*}_{Y|X,n}-P^{(2)}_{Y|X,n}\right\|_{TV}\\
      &\leq\left\|P^{(1)}_{Y|X,n}-P^{*}_{Y|X}\right\|_{TV}+\left\|P^{*}_{Y|X}-P^{(2)}_{Y|X,n}\right\|_{TV}\\
      &\leq2\max\left\{\left\|P^{(1)}_{Y|X,n}-P^{*}_{Y|X}\right\|_{TV},\left\|P^{*}_{Y|X}-P^{(2)}_{Y|X,n}\right\|_{TV}\right\}
    \end{align}
    which leads to 
    \begin{align}
      c<\max\left\{\left\|P^{(1)}_{Y|X,n}-P^{*}_{Y|X,n}\right\|_{TV},\left\|P^{*}_{Y|X,n}-P^{(2)}_{Y|X,n}\right\|_{TV}\right\}.
      \label{eq:27}
    \end{align}
    Again the right hand side is seen to be computable under the assumption that
    $F_{\mathrm{opt}}$ is computable. With this it is now possible to construct
    the characteristic function of the set $\mathcal{A}$ by calculating the
    right hand side of~\eqref{eq:27}. Because $c$ is a rational lower bound and
    the right hand side is computable the comparison is effectively
    decidable~(see~\cite{PourEl2016}).
      
    But with this we have a way to algorithmically compute the
    characteristic function of the set $\mathcal{A}$ thus making the set
    effectively decidable. This shows that the assumption that
    $F_{\mathrm{opt}}$ is computable cannot be true. Because we took any
    function in $\mathcal{M}(\mathcal{X},\mathcal{Y},P_{X})$ as
    $F_{\mathrm{opt}}$ no function in this set is Banach-Mazur computable.
  \end{proof}
\end{lemma}
So far we have already shown that the general rate distortion problem cannot be
solved by a Turing machine or some constructive mathematical algorithm based on
recursive functions. 

\begin{figure}[ht]
  \centering
  \includegraphics[width=.6\textwidth]{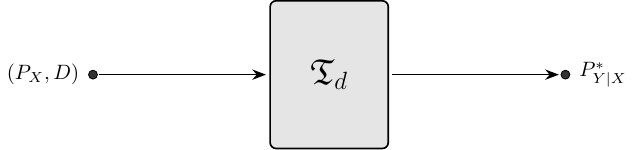}
  \caption{Block diagram of a Turing machine computing $P^{*}_{Y|X}$. Now the
    Turing machine only gets a representation of $P_{X}$ and $D$ as Input.  Even
    if there exist a Turing machine of this type for every $d$ there does not
    necessarily exist a Turing machine of the more general kind shown
    in figure~\protect\ref{fig:turing-picto1}.}\label{fig:turing-picto3}
\end{figure}

Nevertheless this raises the natural question whether the problem is
constructively solvable at least for some fixed distortion measures
$d:\mathcal{X}\times\mathcal{Y}\to\mathbb{R}_{\geq0, \mathrm{c}}$ if only the
source probability and distortion level are used as input for
$F_{\mathrm{opt}}$. This is equivalent to the idea of finding a function
$F_{\mathrm{opt}}$ in the smaller set $\mathcal{M}(\mathcal {X},
\mathcal{Y},d)$. A Turing machine for this case is shown in
figure~\ref{fig:turing-picto3}.

In contrast to the last theorem we now are interested in the set of functions
\(F_{\mathrm{opt}}\in\mathcal{M}(\mathcal{X}, \mathcal{Y}, d)\). Here the
distortion measure as well as the alphabets are fixed and only a computable
source probability $P_{X}\in\mathcal{P}_{\mathrm{c}}(\mathcal{X})$ as well as a
computable distortion level $D\in\mathbb{R}_{\geq0,\mathrm{c}}$ are used as
argument for $F_{\mathrm{opt}}$.

As it turns out while the problem is in general better behaved as in the former
case the problem is not Banach-Mazur computable even under very mild conditions
on the now fixed distortion measures $d$. These results even holds for the
important and well known Hamming distortion measure.

\begin{theorem}
  Let a computable distortion measure
  \(d:\mathcal{X}\times\mathcal{Y}\to\mathbb{R}_{\geq0,\mathrm{c}}\) be
  given. Let the source $|\mathcal{X}|\geq2$ and reconstruction $|\mathcal{Y}|\geq2$
  alphabets be arbitrary but finite. 

  Then under the assumption that $R(D)$ is not equivalent to the zero function for
  every input distribution there exist no function
  $F_{\mathrm{opt}}\in\mathcal{M}(\mathcal{X}, \mathcal{Y}, d)$ that is
  Banach-Mazur and therefore Turing computable.
\end{theorem}

Again we start by analyzing any of the functions
$F_{\mathrm{opt}}\in\mathcal{M}(\mathcal{X}, \mathcal{Y}, d)$ the result then
follows from the following theorems.

We first give two easy examples to get a better understanding of the underlying
problem. These are then extended to the above theorem later on.
\begin{lemma}
  \begin{figure}[ht!]
    \centering
    \includegraphics[width=.3\textwidth]{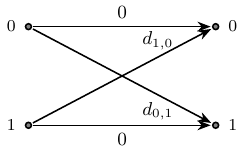}
    \caption{Hamming Distortion Measure with $|\mathcal{X}|=|\mathcal{Y}|=2$.}
  \end{figure}
  
  Let a single letter distortion measure \(d:\mathcal{X}\times\mathcal{Y}\to\mathbb{R}_{\mathrm{c}}\) with
  \(|\mathcal{X}|=2\), \(|\mathcal{Y}|=2\) and $d_{0,1}>0$, $d_{1,0}>0$ be given.
  \begin{equation}
    d= 
    \begin{bmatrix} 
      0       & d_{0,1} \\
      d_{1,0} & 0
    \end{bmatrix}
  \end{equation}
  Then there exist a computable sequence of input distributions
  \({(P_{X,n})}_{n\in\mathbb{N}}\in\mathcal{P}_{\mathrm{c}}(\mathcal{X})\) and a
  computable maxmial distortion level $D_{\max}\in\mathbb{R}_{\mathrm{c}}$ such
  that no $F_{\mathrm{opt}}$ computing a sequence of optimal transition probability matrices $P_{Y|X,n}^{*}$ is Banach-Mazur computable.
  
  \begin{proof}
    We begin by setting \(P_{X}(0)=\frac{d_{1,0}}{d_{0,1}+d_{1,0}}\) as well as
    \(P_{X}(1)=\frac{d_{0,1}}{d_{0,1}+d_{1,0}}\).  For this input distribution
    $D_{\max}$ is easily found to be given by
    \begin{equation}
      D_{\max}=\frac{d_{1,0}d_{0,1}}{d_{0,1}+d_{1,0}}
    \end{equation}
    This can be seen to be correct by only considering one output $P_{Y}(0)$ or
    $P_{Y}(1)$.
    
    Now let $\mathcal{A}$ be a recursively enumerable non recursive set. Then
    there exist a recursive function $a$ with
    \(\mathrm{range(a)}=\mathcal{A}\). To ensure that the sequence of probability
    vectors is non-negative select a $\tilde{m}$ such that
    \(\min\left(P_{X}(0),P_{X}(1)\right)\geq 2^{-\tilde{m}}\).
    
    We then define sequence in terms of $a$ by
    \begin{equation}
      x_{n,k}= 
      \begin{cases}
        2^{-m},& \text{if}\quad n=a(m),\ \tilde{m}\leq m\leq k\\ 
        0&\text{otherwise.}
      \end{cases}
    \end{equation}
    With this we define two sequences of source probability distributions
    \begin{equation}
      P^{(1)}_{X,n,k}=
      \begin{pmatrix}
        P_{X}(0)+x_{n,k}\\
        P_{X}(1)-x_{n,k}
      \end{pmatrix}
    \end{equation}
    and
    \begin{equation}
      P^{(2)}_{X,n,k}=
      \begin{pmatrix}
        P_{X}(0)-x_{n,k}\\
        P_{X}(1)+x_{n,k}
      \end{pmatrix}
    \end{equation}
    
    \begin{equation}
      P^{(1)}_{X,n}=\lim_{k\to\infty}P^{(1)}_{X,n,k}
    \end{equation}
    
    \begin{equation}
      P^{(2)}_{X,n}=\lim_{k\to\infty}P^{(2)}_{X,n,k}
    \end{equation}
    
    By the consideration of the total variation distance for \(k\geq K+1\)
    \begin{align}
      \| P^{(1)}_{X,n}- P^{(1)}_{X,n,k}\|_{l_1}&=\| P^{(2)}_{X,n}- P^{(2)}_{X,n,k}\|_{l_1}\\
      &=2\|x_{n}-x_{n,k}\|_{l_1}\leq\frac{1}{2^{K}}
    \end{align}
    the convergence of these sequences in $k$ is effective for every \(n\in\mathbb{N}\).
    
    Now consider the optimal transition probabilities for these sequences
    \(P^{(1)}_{X,n}\),\(P^{(2)}_{X,n}\) for the case $n\in\mathcal{A}$.
    \begin{equation}
      P^{(1)}_{Y|X}=
      \begin{pmatrix}
        1 & 0 \\
        1 & 0 \\
      \end{pmatrix}
    \end{equation}
    This is because coding only the slighter greater of the two probabilities will
    lead to a smaller distortion with the same rate $R(D)=0$.
    
    For the other sequence we find in the same way
    \begin{equation}
      P^{(2)}_{Y|X,*}=
      \begin{pmatrix}
        0 & 1 \\
        0 & 1 \\
      \end{pmatrix}.
    \end{equation}
    
    If $n\notin\mathcal{A}$ we find that both transition probabilities are
    possible solutions to the problem.
    
    Because of the convexity of the rate distortion problem the optimal transition
    probabilities are then given by all convex combinations of those two solutions
    \begin{equation}
      P^{*}_{Y|X}\in\left\{\lambda
      \begin{pmatrix} 
        1 & 0 \\
        1 & 0
      \end{pmatrix}
      +(1-\lambda)
      \begin{pmatrix} 
        0 & 1 \\
        0 & 1
      \end{pmatrix}
      \Bigg|\lambda\in[0,1].
      \right\}
    \end{equation}
    
    Further we get
    \begin{equation}
      \left\|P^{(1)}_{Y|X}-P^{(2)}_{Y|X}\right\|_{TV}=\frac{1}{2}\max_{x\in\mathcal{X}}\sum_{y\in\mathcal{Y}}\left|P^{(1)}_{Y|X}(y|x)-P^{(2)}_{Y|X}(y|x)\right|=1
    \end{equation}
    for the total variation distance of these two optimal transistion probabilities.
    
    By using the above solutions for the transition probabilities we get
    \begin{align}
      1=\left\|P^{(1)}_{Y|X}-P^{(2)}_{Y|X}\right\|_{TV}&=\left\|P^{(1)}_{Y|X}+P^{*}_{Y|X}-P^{*}_{Y|X}-P^{(2)}_{Y|X}\right\|_{TV}\\
      &\leq\left\|P^{(1)}_{Y|X}-P^{*}_{Y|X}\right\|_{TV}+\left\|P^{*}_{Y|X}-P^{(2)}_{Y|X}\right\|_{TV}\\
      &\leq2\max\left\{\left\|P^{(1)}_{Y|X}-P^{*}_{Y|X}\right\|_{TV},\left\|P^{*}_{Y|X}-P^{(2)}_{Y|X}\right\|_{TV}\right\}
    \end{align}

    \begin{align}
      \frac{1}{2}\leq\max\left\{\left\|P^{(1)}_{Y|X,*}-P^{*}_{Y|X}\right\|_{TV},\left\|P^{*}_{Y|X}-P^{(2)}_{Y|X,*}\right\|_{TV}\right\}
    \end{align}
  \end{proof}
\end{lemma}

\begin{lemma}
  Let the single letter distortion measure
  \(d:\mathcal{X}\times\mathcal{Y}\to\mathbb{R}_{\mathrm{c}}\) from the previous theorem with
  \(|\mathcal{X}|=2\), \(|\mathcal{Y}|=2\) and $d_{0,1}>0$, $d_{1,0}>0$ be given.
  \begin{equation}
    d= 
    \begin{bmatrix} 
      0      & d_{0,1} \\
      d_{1,0} & 0
    \end{bmatrix}
  \end{equation}
  Then there exist sequence of computable input distributions
  \({(P_{X,n})}_{n\in\mathbb{N}}\in\mathcal{P}_{\mathrm{c}}(\mathcal{X})\) and a
  computable sequence of distortion levels $D_{n}$ such that no function
  $F_{\mathrm{opt}}$ computing the sequence of optimal transition probabilities
  $P_{Y|X,n}^{*}$ in this case is a Banach-Mazur computable function.
  
  \begin{proof}
    We start by giving the optimal transition probabilities for the input sequence
    \begin{equation}
      P_{X}=
      \begin{pmatrix}
        1 \\ 0
      \end{pmatrix}
    \end{equation}
    and the distortion level $D=0$.  In this case we are not allowed to tolerate any
    errors. Nevertheless because \(P_{X}(1)=0\) the coding for the second input letter is
    irrelevant. This shows that the optimal transition probabilities are given by the set
    \begin{equation}
      P^{*}_{Y|X}\in\left\{\lambda
      \begin{pmatrix} 
        1 & 0 \\
        1 & 0
      \end{pmatrix}
      +(1-\lambda)
      \begin{pmatrix} 
        1 & 0 \\
        0 & 1
      \end{pmatrix}
      \Bigg|\lambda\in[0,1].
      \right\}.
    \end{equation}
    
    Now let $\mathcal{A}$ be a recursively enumerable non recursive set. Then there exist a recursive
    function $a$ with \(\mathrm{range(a)}=\mathcal{A}\).  Further select a $\tilde{m}$ such that
    \(\frac{d_{0,1}d_{1,0}}{d_{0,1}+d_{1,0}}> 2^{-\tilde{m}}\).  We then define the sequence
    \begin{equation}
      x_{n,k}= 
      \begin{cases}
        2^{-m},& \text{if}\quad n=a(m),\ \tilde{m}\leq m\leq k\\ 
        0&\text{otherwise.}
      \end{cases}
    \end{equation}
    
    Further we select a $L$ such that
    \begin{equation}
      D_{n,k}=x_{n,k}\frac{d_{1,0}}{2L} \leq x_{n,k}
    \end{equation}
    
    We then again define two input probabilities by
    \begin{equation}
      P^{(1)}_{X,n,k}=
      \begin{pmatrix}
        1-x_{n,k}\\
        x_{n,k}
      \end{pmatrix}
    \end{equation}
    and
    \begin{equation}
      P^{(2)}_{X,n,k}=
      \begin{pmatrix}
        1-\frac{x_{n,k}}{2L}\\
        \frac{x_{n,k}}{2L}
      \end{pmatrix}
    \end{equation}
    
    By the same argumentation as in the other proof all these sequences converge effectively to 
    \begin{equation}
      P^{(1)}_{X,n}=\lim_{k\to\infty}P^{(1)}_{X,n,k}
    \end{equation}
    
    \begin{equation}
      P^{(2)}_{X,n}=\lim_{k\to\infty}P^{(2)}_{X,n,k}
    \end{equation}
    and
    \begin{equation}
      D_{n}=\lim_{k\to\infty}D_{n,k}
    \end{equation}
    
    We then find that the first sequence \(P^{(1)}_{X,n}\) has the following lower bound on the distortion
    for \(x_{n}\neq0\)
    \begin{equation}
      D_{n}\geq P^{(1)}_{X,n}(1)d_{1,0}(1-P^{(1)}_{Y|X}(1|1))
    \end{equation}
    in terms of the optimizing transition probability $P_{Y|X}$
    
    From this we then get a lower bound
    \begin{equation}
      P^{(1)}_{Y|X}(1|1)\geq\frac{D_{n}}{x_{n}d_{1,0}}\geq\frac{1}{2}
    \end{equation}
    for the optimal transition probability.
    
    In case of $n\notin\mathcal{A}$ we get \(x_{n}=0\) and we can use the same bound for the optimal 
    transition probability $P^{*}_{Y|X}$ because then we have $P^{(1)}_{X,n}(1)=0$ and the coding of the letter $1$
    doesn't change the rate distortion function.
    
    For the second input distribution we find
    \begin{equation}
      D^{(2)}_{\max,n}=P^{(2)}_{X,n}(1)d_{1,0}=x_{n}\frac{d_{1,0}}{2L}
    \end{equation}
    in this case the optimal transition probability only codes the first input letter and we get $R(D_{\max})=0$.
    \begin{equation}
      P^{(2)}_{Y|X}=
      \begin{pmatrix}
        1 & 0\\
        1 & 0\\
      \end{pmatrix}
    \end{equation}
    
    Now by comparing both solutions we get the following bound for the total variation distance between
    \begin{align}
      \left\|P^{(1)}_{Y|X}-P^{(2)}_{Y|X}\right\|_{TV}&=\frac{1}{2}\max_{x\in\mathcal{X}}\sum_{y\in\mathcal{Y}}\left|P^{(1)}_{Y|X}(y|x)-P^{(2)}_{Y|X}(y|x)\right|\\
      &\geq\frac{1}{2}\left(\left|1-\frac{1}{2}\right|+\left|\frac{1}{2}-0\right|\right)\\
      &\geq\frac{1}{2}.
    \end{align}
    the two transition probabilities. 
    
    With this we have together with $P^{*}_{Y|X}$
    \begin{align}
      \frac{1}{2}=\left\|P^{(1)}_{Y|X}-P^{(2)}_{Y|X}\right\|_{TV}&=\left\|P^{(1)}_{Y|X}+P^{*}_{Y|X}-P^{*}_{Y|X}-P^{(2)}_{Y|X}\right\|_{TV}\\
      &\leq\left\|P^{(1)}_{Y|X}-P^{*}_{Y|X}\right\|_{TV}+\left\|P^{*}_{Y|X}-P^{(2)}_{Y|X}\right\|_{TV}\\
      &\leq2\max\left\{\left\|P^{(1)}_{Y|X}-P^{*}_{Y|X}\right\|_{TV},\left\|P^{*}_{Y|X}-P^{(2)}_{Y|X}\right\|_{TV}\right\}
    \end{align}
    and therefore
    \begin{align}
      \frac{1}{4}\leq\max\left\{\left\|P^{(1)}_{Y|X,*}-P^{*}_{Y|X}\right\|_{TV},\left\|P^{*}_{Y|X}-P^{(2)}_{Y|X,*}\right\|_{TV}\right\}.
    \end{align}
    From here on the proof is the same as in the previous case.
  \end{proof}
\end{lemma}

In the following we will extend the previous results to arbitrary dimensions and distortion
measures.  We begin by the following simple lemma regarding the structure of an arbitrary normal
distortion measure.

\begin{lemma}%
  \label{lem:distortion-matrix}
  Let $\mathcal{X}$ be a finite source and $\mathcal{Y}$ be a finite
  reproduction alphabet with \(|\mathcal{X}|\geq2\) and $|\mathcal{Y}|\geq2$ and
  let $d:\mathcal{X}\times\mathcal{Y}\to\mathbb{R}_{\geq0}$ be any
  normal distortion measure.
  
  Further assume that there exist at least one source distribution $P_{X}$ such that the rate
  distortion function $R(D)$ is not equivalent to the zero function.

  Then there exist at least two distinct letters in the source alphabet
  $k_{1},k_{2}\in\mathcal{X}$ and two distinct letters in the reproduction alphabet
  $l_{1},l_{2}\in\mathcal{Y}$ such that $d(k_{1},l_{1})=0$ but $d(k_{1},l_{2})>0$ as well as
  $d(k_{2},l_{2})=0$ but $d(k_{2},l_{1})>0$ hold.
  
  \begin{proof}
    Because $d$ is a normal distortion measure for every \(x\in\mathcal{X}\) there exist at
    least one \(y\in\mathcal{Y}\) such that \(d(x,y)=0\).  Further we can assume that for every
    $l$ there exist at least one $k$ with \(d(k,l)>0\) as otherwise we always have $R(D)=0$ for
    all $D$ and $P_{X}$ by simply coding only this $l$.
    
    We start by selecting any source letter $k_{1}$ we then have \(d(k_{1},l_{1})=0\) for some
    $l_{1}$ because $d$ is normal and there exist at least one source letter $k_{2}$ such that
    $d(k_{2},l_{1})>0$ for this $l_{1}$. We also know that we have $d(k_{2},l_{2})=0$ for this
    $k_{2}$ and some $l_{2}$. Now, if for this $l_{2}$ we also have $d(k_{1},l_{2})>0$ we are
    already done. This is always the case if there are only two source letters.
    
    Now assume that we also have $d(k_{1},l_{2})=0$. But then again there exist another $k_{3}$
    such that $d(k_{3},l_{2})>0$ and some $l_{3}$ such that $d(k_{3},l_{3})=0$.
    Note that we have $d(k_{3},l_{1})>0$ as otherwise we are again finished by replacing $k_{2}$
    with $k_{3}$ in the above equations.
    
    With this we then have either that $d(k_{2},l_{3})>0$ and $d(k_{2},l_{2})=0$ or
    $d(k_{2},l_{3})=0$ and $d(k_{2},l_{1})>0$ so in every case we found the desired points by
    the pigeonhole principle.
  \end{proof}
\end{lemma}

\begin{lemma}
  Let a computable distortion measure
  \(d:\mathcal{X}\times\mathcal{Y}\to\mathbb{R}_{\geq0,\mathrm{c}}\) be
  given. Moreover assume that the source alphabet as well as the reproduction
  alphabet are arbitrary but finite with $|\mathcal{X}|\geq2$ and
  $|\mathcal{Y}|\geq2$ respectively. Further assume $R(D)$ is not equivalent to
  the zero function for every input distribution.  Then there exist computable
  sequences of input distributions
  $P_{X,n}\in\mathcal{P}_{\mathrm{c}}(\mathcal{X})$ and computable sequences of
  distortion levels $D_{n}=D_{n,\max}$ as well as $D_{n}\to 0$ such that there
  exist no function $F_{\mathrm{opt}}$ in
  $\mathcal{M}(\mathcal{X},\mathcal{Y},d)$ that is Banach-Mazur and therefore
  Turing computable.

  Moreover if in addition to the points in Lemma~\ref{lem:distortion-matrix}
  there exist points \(k_{3},k_{4}\in\mathcal{X}\) such that $d(k_{3},l_{1})>0$
  as well as $d(k_{4},l_{2})>0$ then the sequence $D_{n}=D_{n,\max}$ can be
  chosen as a computable constant.
  \begin{proof}
    We begin by construction of a starting source probability $P_{X}$ needed for
    the proof. For this select an arbitrary rational probability vector as
    starting distribution for $P_{X}$. From Lemma~\ref{lem:distortion-matrix} we
    know that there exist two distinct letters in the input alphabet
    $k_{1},k_{2}\in\mathcal{X}$ and two letter in the outptut alphabet
    $l_{1},l_{2}\in\mathcal{Y}$ such that $d(k_{1},l_{1})=0$ but
    $d(k_{1},l_{2})>0$ as well as $d(k_{2},l_{2})=0$ but $d(k_{2},l_{1})>0$
    hold.
    
    Now we start with the letter $k_{1}$. This letter is correctly decoded as
    $l_{1}$ while there exist at least one other letter $k$ such that an error
    is made whenever we decode $k$ as $l_{1}$. With this we change $P_{X}$ in a
    first step in such a way that we get
    \begin{equation}
      \sum_{k=0}^{K-1}P_{X}(k)d(k,l_{1})<\sum_{k=0}^{K-1}P_{X}(k)d(k,l),\qquad\text{for all}\quad l
      \label{eq:28}.
    \end{equation}
    This is always possible by increasing the probability for $P_{X}(k_{1})$ and
    decreasing the probability for any of the other $k$ for which $d(k,l_{1})>0$
    holds.
    
    With this $l_{1}$ we then want to further change $P_{X}$ such that we also
    have
    \begin{equation}
      \sum_{k=0}^{K-1}P_{X}(k)d(k,l_{1})=\sum_{k=0}^{K-1}P_{X}(k)d(k,l_{2})
      \label{eq:29}
    \end{equation}
    for $l_{2}$ and~\eqref{eq:28} is accordingly satisfied for both $l_{1}$ and $l_{2}$.
    
    By use of equation~\eqref{eq:29} and the properties of a probability vector we find that this
    can be achieved by setting
    \begin{equation}
      P_{X}(k_{2})=\frac{d(k_{1},l_{2})}{d(k_{1},l_{2})+d(k_{2},l_{1})}\left(1-\sum_{\substack{k=0\\k\notin \{k_{1},k_{2}\}}}^{K-1}P_{X}(k)\right)+
      \sum_{\substack{k=0\\k\notin \{k_{1},k_{2}\}}}^{K-1}P_{X}(k)\frac{d(k,l_{2})-d(k,l_{1})}{d(k_{1},l_{2})+d(k_{2},l_{1})}.
    \end{equation}
    again by changing all other $P_{X}$ appropriately. 
    
    We further proceed by defining the necessary sequences needed to show that
    $F_{\mathrm{opt}}$ cannot be a computable function.  For this let
    $\mathcal{A}$ be a recursively enumerable non-recursive set. Then there
    exist a recursive function $a(m)$ with
    \(\mathrm{range}(a)=\mathcal{A}\). Further let $\tilde{m}$ be the biggest
    positive integer such that
    \begin{equation}
      \frac{\min_{k}P_{X}(k)}{\max_{k,l}d(k,l)}\geq 2^{-\tilde{m}}.
    \end{equation}
    We further define
    \begin{equation}
      x_{n,m}:= 
      \begin{cases}
        2^{-i},& \text{if}\quad n=a(i),\text{ for } \tilde{m}<i\leq m\\ 
        0&\text{otherwise.}
      \end{cases}.
    \end{equation}
    
    We proceed by defining the following sequences of source probabilities based on the
    constructed distribution $P_{X}$
    \begin{equation}
      P^{(1)}_{X,n,m}(k):=
      \begin{cases}
        P_{X}(k_{1})+x_{n,m}& k=k_{1}\\
        
        P_{X}(k_{2})-x_{n,m}& k=k_{2}\\
        P_{X}(k)&\text{otherwise}.
      \end{cases}
    \end{equation}
    as well as
    \begin{equation}
      P^{(2)}_{X,n,m}(k):=
      \begin{cases}
        P_{X}(k_{1})-\frac{d(k_{2},l_{1})}{d(k_{1},l_{2})}x_{n,m}& k=k_{1}\\
        
        P_{X}(k_{2})+x_{n,m}& k=k_{2}\\
        P_{X}(k)&\text{otherwise}.
      \end{cases}
    \end{equation}
    Again it can be easily shown that
    \begin{equation}
      P^{(1)}_{X,n}(k)=\lim_{m\to\infty}P^{(1)}_{X,n,m}(k)
    \end{equation}
    and
    \begin{equation}
      P^{(2)}_{X,n}(k)=\lim_{m\to\infty}P^{(1)}_{X,n,m}(k)
    \end{equation}
    converge and the convergence is also effective. So both input sequences are sequences of
    computable real numbers. 
    
    We want to calculate the optimal transition probabilities for both sequences
    at $D_{\max,n}$.  Note that because of the construction of $P_{X}$ and
    therefore also the construction of $P^{(1)}_{X,n}$ and $P^{(2)}_{X,n}$ it
    is optimal to code only $l_{1}$ or $l_{2}$ at $D_{\max,n}$ respectively.
    Moreover we find
    \begin{align}
      D_{\max,n,m}&=\sum_{k\in\mathcal{X}}P^{(1)}_{X,n}(k)d(k,l_{1})=\sum_{k\in\mathcal{X}}P^{(2)}_{X,n}(k)d(k,l_{2})\\
      &=\sum_{k\in\mathcal{X}}P_{X}(k)d(k,l_{1})-x_{n,m}d(k_{2},l_{1})\\
      &=\sum_{k\in\mathcal{X}}P_{X}(k)d(k,l_{2})-x_{n,m}d(k_{2},l_{1})
    \end{align}
    so that $D_{\max,n}$ is the same when coding $l_{1}$ or $l_{2}$.
    
    Employing the usual technique we find that also this convergence is effective and we have
    \begin{equation}
      D_{\max,n}=\lim_{m\to\infty}D_{\max,n,m}.
    \end{equation}
    With this we analyze the optimal transition probabilities for
    $P^{(1)}_{X,n}$ and $P^{(2)}_{X,n}$ at $D_{\max,n}$.
    
    We start with $P^{(1)}_{X,n}$. If $n\in\mathcal{A}$ then we get the minimal distortion by simply
    coding only $l_{1}$ as for this we get the minimal distortion by~\eqref{eq:28}.
    The optimal transition probability in this case is therefore given by
    \begin{equation}
      P^{(1)}_{Y|X}(y|x)=
      \begin{cases}
        1 & y=l_{1}\\
        0 & \text{otherwise}
      \end{cases}
    \end{equation}
    For $P^{(2)}_{X,n}$ we get the same result for $n\in\mathcal{A}$ but this
    time it is best to only code $l_{2}$.
    \begin{equation}
      P^{(2)}_{Y|X}(y|x)=
      \begin{cases}
        1 & y=l_{2}\\
        0 & \text{otherwise}
      \end{cases}
    \end{equation}
    
    In case $n\notin\mathcal{A}$ coding only $l_{1}$ as well as coding only
    $l_{2}$ will result in a minimal $D_{\max}$. So for the optimal transition
    probability we get therefore that every convex combination of
    \(P^{(1)}_{Y|X}(y|x)\) and \(P^{(2)}_{Y|X}(y|x)\) are solutions and the
    optimal transition probabilities $P^{*}_{Y|X}$ for both cases are given by
    the set
    \begin{equation}
      \left\{P^{*}_{Y|X}=\lambda P^{(1)}_{Y|X}(y|x)+(1-\lambda)P^{(2)}_{Y|X}(y|x)\Bigg|\lambda\in[0,1]\right\}.
    \end{equation}
    Finally we get the following
    \begin{equation}
      \left\|P^{(1)}_{Y|X}-P^{(2)}_{Y|X}\right\|_{TV}=\frac{1}{2}\max_{x\in\mathcal{X}}\sum_{y\in\mathcal{Y}}\left|P^{(1)}_{Y|X}(y|x)-P^{(2)}_{Y|X}(y|x)\right|=1
    \end{equation}
    and further
    \begin{align}
      1=\left\|P^{(1)}_{Y|X}-P^{(2)}_{Y|X}\right\|_{TV}&=\left\|P^{(1)}_{Y|X}+P^{*}_{Y|X}-P^{*}_{Y|X}-P^{(2)}_{Y|X}\right\|_{TV}\\
      &\leq\left\|P^{(1)}_{Y|X}-P^{*}_{Y|X}\right\|_{TV}+\left\|P^{*}_{Y|X}-P^{(2)}_{Y|X}\right\|_{TV}\\
      &\leq2\max\left\{\left\|P^{(1)}_{Y|X}-P^{*}_{Y|X}\right\|_{TV},\left\|P^{*}_{Y|X}-P^{(2)}_{Y|X}\right\|_{TV}\right\}
    \end{align}
    Finally we have
    \begin{align}
      \frac{1}{2}\leq\max\left\{\left\|P^{(1)}_{Y|X}-P^{*}_{Y|X}\right\|_{TV},\left\|P^{*}_{Y|X}-P^{(2)}_{Y|X}\right\|_{TV}\right\}.
    \end{align}
    and because we can effectively decide whether any computable number is \(1/2\)
    or $0$ we find by the same argumentation as in the proof above that
    $F_{\mathrm{opt}}$ is not a computable function of $D_{\max,n}$ and $P_{X,n}$.
    
    To finish the proof we only have to show that $P^{(1)}_{X,n}$,$P^{(2)}_{X,n}$
    can be chosen in such a way that $D_{\max,n}$ is independent of $n$ if there
    exist additional points $k_{3}$ and $k_{4}$.  This follows easily by regarding
    the construction and changing the added $x_{n,m}$ terms in $P^{(1)}_{X,n}$ and
    $P^{(2)}_{X,n}$ such that they cancel out only for one $l_{1}$ or $l_{2}$.
    The case $D_{n}\to0$ follows in a similar way.
  \end{proof}
\end{lemma}

From the the above theorems and their proofs it also follows similar as
in~\cite{Boche2023} that we cannon even approximate the $F_{\mathrm{opt}}$
in the above cases.
This is because if we could approximate $F_{\mathrm{opt}}$ to any possible
error we would be able to decide the non decidable sets in the proof.

\section{Conclusion}
Rate distortion theory answers the question of achievable source coding rates
given a distortion measure and a variable distortion level as fidelity
criterion.

The results of rate distortion theory are of high importance in understanding
lossy source coding of continuous sources, joint source channel coding and
compressed sensing for example. The calculation of the rate distortion function
and optimizing test channel probability thus are important tasks in information
theory.

As there are now general analytical solutions to this problem algorithms like
the Blahut-Ariomoto type algorithms are commonly employed, extended and used to
calculate the rate distortion function and to approximate optimizing conditional
probability distributions. Convergence of these algorithms to the optimizer with
a suitable error criterion have been shown only in some special cases.

In this paper we have shown that similar to the behavior in other information
theoretic problems there cannot exist a universal algorithm to compute the
optimizer for rate distortion function. Moreover even in the case of a Hamming
distortion measure there exist, not even a function approximating the optimal
test channel probability, for all distortion levels.

\printbibliography%

@Book{Gallager1968,
  author    = {Robert G. Gallager},
  publisher = {John Wiley \& Sons},
  title     = {Information Theory and Reliable Communication},
  year      = {1968},
}

@Book{Csiszar2011,
  author    = {Imre Csiszár AND János Körner},
  publisher = {Cambridge University Press},
  title     = {Information Theory: Coding Theorems for Discrete Memoryless Systems},
  year      = {2011},
  edition   = {2},
  doi       = {10.1017/cbo9780511921889},
}

@Book{Yeung2008,
  author    = {Raymond W. Yeung},
  publisher = {Springer},
  title     = {Information Theory and Network Coding},
  year      = {2008},
  isbn      = {978-0-387-79233-0},
  series    = {Information Technology: Transmission, Processing, and Storage},
}

@Book{Berger1971,
  author    = {Toby Berger},
  editor    = {Thomas Kailath},
  publisher = {Prentice-Hall},
  title     = {Rate Distortion Theory: A Mathematical Basis for Data Compression},
  year      = {1971},
  address   = {Englewood Cliffs, New Jersey},
  series    = {Information and System Sciences},
}

@Book{Manin2010,
  author    = {Yu. I. Manin},
  editor    = {S. Axler AND K. A. Ribet},
  publisher = {Springer},
  title     = {A Course in Mathematical Logic for Mathematicians},
  year      = {2010},
  edition   = {Second},
  number    = {53},
  series    = {Graduate Text in Mathematics},
  doi       = {DOI 10.1007/978-1-4419-0615-1},
}

@Book{Marshall2011,
  author    = {Albert W. Marshall AND Ingram Olkin AND Barry C. Arnold},
  publisher = {Springer},
  title     = {Inequalities: Theory of Majorization and its Applications},
  year      = {2011},
  edition   = {Second},
  series    = {Springer Series in Statistics},
  doi       = {DOI 10.1007/978-0-387-68276-1},
}

@Book{PourEl2016,
  author    = {Marian B. Pour-El and J. Ian Richards},
  editor    = {Arnold Beckmann},
  publisher = {Cambridge University Press},
  title     = {Computability in Analysis and Physics},
  year      = {2016},
  edition   = {Second},
  series    = {Perspectives in Mathematical Logic},
  doi       = {https://doi.org/10.1017/9781316717325},
}

@Book{Weihrauch2000,
  author    = {Klaus Weihrauch},
  editor    = {W. Brauer and G. Rozenberg and A. Salomaa},
  publisher = {Springer},
  title     = {Computable Analysis: An Introduction},
  year      = {2000},
  isbn      = {ISBN 3540668179},
  series    = {Text in Theoretical Computer Science},
}

@Book{Rogers1967,
  author    = {Hartley Rogers},
  editor    = {E. H. Spanier},
  publisher = {McGraw-Hill},
  title     = {Theory of Recursive Functions and Effective Computability},
  year      = {1967},
  series    = {McGraw-Hill Series in Higher Mathematics},
}

@Article{Arimoto1972,
  author  = {Suguru Arimoto},
  journal = {IEEE Transactions of Information Theory},
  title   = {An Algorithm for Computing the Capacity of Arbitrary Discrete Memoryless Channels},
  year    = {1972},
  month   = jan,
  number  = {1},
  pages   = {14 -- 20},
  volume  = {18},
}

@Article{Erokhin1958,
  author  = {V. Erokhin},
  journal = {Teor.Veroyatnost. i Primenen},
  title   = {e-Entropy of a Discrete Random Variable},
  year    = {1958},
  number  = {1},
  pages   = {103--107},
  volume  = {3},
}

@Article{Boukris1973,
  author  = {P. Boukris},
  journal = {IEEE Transactions on Information Theory},
  title   = {An Upper Bound on the Speed of Convergence of the Blahut Algorithm for Computing Rate-Distortion Functions},
  year    = {1973},
  month   = sep,
  pages   = {708--709},
}

@Article{Csiszar1974,
  author  = {Imre Csiszár},
  journal = {IEEE Transactions on Information Theory},
  title   = {On the Computation of Rate-Distortion Functions},
  year    = {1974},
  month   = jan,
}

@Article{Kostina2012,
  author  = {Victoria Kostina AND Sergio Verdú},
  journal = {IEEE Transactions on Information Theory},
  title   = {Fixed-Length Lossy Compression in the Finite Blocklength Regime},
  year    = {2012},
  month   = jun,
  number  = {6},
  pages   = {3309--3338},
  volume  = {58},
}

@Article{Kostina2014,
  author  = {Victoria Kostina AND Yury Polyanskiy AND Sergio Verdú},
  journal = {IEEE International Symposium on Information Theory},
  title   = {Variable-length Compression allowing errors},
  year    = {2014},
}

@Article{Blahut1972,
  author  = {Richard E. Blahut},
  journal = {IEEE Transactions on Information Theory},
  title   = {Computation of Channel Capacity and Rate-Distortion Functions},
  year    = {1972},
  month   = jul,
  number  = {4},
  pages   = {460 --473},
  volume  = {18},
}

@Article{Meister1967,
  author  = {Bernd Meister AND Werner Oettli},
  journal = {Information and Control},
  title   = {On the Capacity of a Discrete, Constant Channel},
  year    = {1967},
  pages   = {341--351},
  volume  = {11},
}

@Article{Csiszar1984,
  author  = {Imre Csiszár and G. Tusnády},
  journal = {Statistics and Decisions},
  title   = {Information Geometry and Alternating Minimization Procedures},
  year    = {1984},
  number  = {1},
  pages   = {205--237},
}

@Article{Boche2023,
  author   = {Boche, Holger and Schaefer, Rafael F. and Poor, H. Vincent},
  journal  = {IEEE Transactions on Information Theory},
  title    = {Algorithmic Computability and Approximability of Capacity-Achieving Input Distributions},
  year     = {2023},
  number   = {9},
  pages    = {5449-5462},
  volume   = {69},
  doi      = {10.1109/TIT.2023.3278705},
}

@Article{Grigorescu2025,
  author   = {Boche, Holger and Grigorescu, Andrea and Schaefer, Rafael F. and Vincent Poor, H.},
  journal  = {IEEE Transactions on Information Theory},
  title    = {Algorithmic Computability of the Capacity of Additive Colored Gaussian Noise Channels},
  year     = 2025,
  number   = 10,
  pages    = {7419-7434},
  volume   = 71,
  doi      = {10.1109/TIT.2025.3594999},
}

@InProceedings{Grigorescu2024,
  author    = {Boche, Holger and Grigorescu, Andrea and Schaefer, Rafael F. and Poor, H. Vincent},
  booktitle = {2024 IEEE International Symposium on Information Theory (ISIT)},
  title     = {On the Non-Computability of Convex Optimization Problems},
  year      = {2024},
  pages     = {3083-3088},
  doi       = {10.1109/ISIT57864.2024.10619549},
}

@Article{Lee2024,
  author   = {Lee, Yunseok and Boche, Holger and Kutyniok, Gitta},
  journal  = {IEEE Transactions on Information Theory},
  title    = {Computability of Optimizers},
  year     = {2024},
  number   = {4},
  pages    = {2967-2983},
  volume   = {70},
  doi      = {10.1109/TIT.2023.3347071},
}

@Article{Grigorescu2024a,
  author   = {Grigorescu, Andrea and Boche, Holger and Schaefer, Rafael F. and Vincent Poor, H.},
  journal  = {IEEE Transactions on Information Theory},
  title    = {Capacity of Finite State Channels With Feedback: Algorithmic and Optimization Theoretic Properties},
  year     = {2024},
  number   = {8},
  pages    = {5413-5426},
  volume   = {70},
  doi      = {10.1109/TIT.2024.3411919},
}

@Article{Fettweis2021,
  author   = {Fettweis, Gerhard P. and Boche, Holger},
  journal  = {IEEE BITS the Information Theory Magazine},
  title    = {{6G}: The Personal Tactile Internet—And Open Questions for Information Theory},
  year     = 2021,
  number   = 1,
  pages    = {71-82},
  volume   = 1,
  doi      = {10.1109/MBITS.2021.3118662},
}

@Article{Specker1959,
  author  = {Ernst Specker},
  journal = {Constructivity in Mathematics},
  title   = {Der Satz vom Maximum in der rekursiven Analysis},
  year    = {1959},
  month   = jan,
  pages   = {254--265},
  volume  = {10},
}

@Article{Boche2020,
  author   = {Boche, Holger and Pohl, Volker},
  journal  = {IEEE Transactions on Information Theory},
  title    = {On the Algorithmic Solvability of Spectral Factorization and Applications},
  year     = {2020},
  number   = {7},
  pages    = {4574-4592},
  volume   = {66},
  doi      = {10.1109/TIT.2020.2968028},
}

@Article{Shannon1948,
  author  = {Shannon, C. E.},
  journal = {The Bell System Technical Journal},
  title   = {A mathematical theory of communication},
  year    = {1948},
  number  = {3},
  pages   = {379-423},
  volume  = {27},
  doi     = {10.1002/j.1538-7305.1948.tb01338.x},
}

@Book{Soare1987,
  author    = {Robert I. Soare},
  publisher = {Springer-Verlag Berlin Heidelberg},
  title     = {Recursively Enumerable Sets and Degrees, A Study of Computable Functions and Computably Generated Sets},
  year      = {1987},
}

@Article{Boche2019,
  author   = {Boche, Holger and Schaefer, Rafael F. and Baur, Sebastian and Poor, H. Vincent},
  journal  = {IEEE Transactions on Signal Processing},
  title    = {On the Algorithmic Computability of the Secret Key and Authentication Capacity Under Channel, Storage, and Privacy Leakage Constraints},
  year     = {2019},
  number   = {17},
  pages    = {4636-4648},
  volume   = {67},
  doi      = {10.1109/TSP.2019.2929467},
}

@Article{Boche2020a,
  author  = {Boche, Holger AND R. F. Schaefer AND H. V. Poor},
  journal = {Communications in Information and Systems},
  title   = {Shannon meets Turing: Non-computability and non approximability of the finite state channel capacity},
  year    = {2020},
  number  = {2},
  pages   = {81--116},
  volume  = {20},
}

@Book{KLeene1971,
  author    = {Stephen Cole Kleene},
  publisher = {North-Holland Publishing Company},
  title     = {Introduction to Metamathematics},
  year      = {1971},
}

@Article{Turing1937,
  author  = {A. M. Turing},
  journal = {Proceedings of the Londong Mathematical Society},
  title   = {On Computable Numbers, with an Application to the Entscheidungsproblem},
  year    = {1937},
  number  = {1},
  pages   = {230--265},
  volume  = {2-42},
  doi     = {https://doi.org/10.1112/plms/s2-42.1.230},
}

@Article{Goedel1930,
  author  = {K. Gödel},
  journal = {Monatshefte für Mathematik und Physik},
  title   = {Die Vollständigkeit der Axiome des logischen Funktionenkalküls},
  year    = {1930},
  pages   = {349--360},
  volume  = {37},
}

@InProceedings{Li2019,
  author    = {Li, Haobo and Cai, Ning},
  booktitle = {2019 IEEE International Symposium on Information Theory (ISIT)},
  title     = {A Blahut-Arimoto Type Algorithm for Computing Classical-Quantum Channel Capacity},
  year      = {2019},
  pages     = {255-259},
  doi       = {10.1109/ISIT.2019.8849608},
}

@InProceedings{Ugur2017,
  author    = {Uğur, Yiğit and Aguerri, Iñaki Estella and Zaidi, Abdellatif},
  booktitle = {2017 IEEE Information Theory Workshop (ITW)},
  title     = {A generalization of blahut-arimoto algorithm to compute rate-distortion regions of multiterminal source coding under logarithmic loss},
  year      = {2017},
  pages     = {349-353},
  doi       = {10.1109/ITW.2017.8277967},
}

@Article{Naiss2013,
  author   = {Naiss, Iddo and Permuter, Haim H.},
  journal  = {IEEE Transactions on Information Theory},
  title    = {Extension of the Blahut–Arimoto Algorithm for Maximizing Directed Information},
  year     = {2013},
  number   = {1},
  pages    = {204-222},
  volume   = {59},
  doi      = {10.1109/TIT.2012.2214202},
}

@InProceedings{Dupuis2004,
  author    = {Dupuis, F. and Yu, W. and Willems, F.M.J.},
  booktitle = {International Symposium onInformation Theory, 2004. ISIT 2004. Proceedings.},
  title     = {Blahut-Arimoto algorithms for computing channel capacity and rate-distortion with side information},
  year      = {2004},
  pages     = {179-},
  doi       = {10.1109/ISIT.2004.1365218},
}

@Book{Cover2006,
  author    = {Thomas M. Cover AND Joy A. Thomas},
  publisher = {John Wiley \& Sons, Inc.},
  title     = {Elements of Information Theory},
  year      = {2006},
}

@Article{Vontobel2008,
  author   = {Vontobel, Pascal O. and Kavcic, Aleksandar and Arnold, Dieter M. and Loeliger, Hans-Andrea},
  journal  = {IEEE Transactions on Information Theory},
  title    = {A Generalization of the Blahut–Arimoto Algorithm to Finite-State Channels},
  year     = {2008},
  number   = {5},
  pages    = {1887-1918},
  volume   = {54},
  doi      = {10.1109/TIT.2008.920243},
}

@InProceedings{Stylianou2024,
  author    = {Stylianou, Evagoras and Charalambous, Charalambos D. and Charalambous, Themistoklis},
  booktitle = {2024 IEEE International Symposium on Information Theory (ISIT)},
  title     = {Implicit and Explicit Formulas of the Joint RDF for a Tuple of Multivariate Gaussian Sources with Individual Square-Error Distortions},
  year      = {2024},
  pages     = {1688-1693},
  doi       = {10.1109/ISIT57864.2024.10619678},
}

@Article{Jorswieck2007,
  author  = {Eduard Jorswieck AND Holger Boche},
  journal = {Foundations and Trends in Communications and Information Theory},
  title   = {Majorization and Matrix-Monotone Functions in Wireless Communications},
  year    = {2007},
  month   = jul,
  number  = {6},
  pages   = {553--701},
  volume  = {3},
}
\end{document}